\documentstyle[psfig,epsfig,times,amssymb,amsmath,fleqn]{mn}

\newif\ifAMStwofonts

\def\simlt{\lower.5ex\hbox{$\; \buildrel < \over \sim \;$}}
\def\simgt{\lower.5ex\hbox{$\; \buildrel > \over \sim \;$}}
\def\etal{{\it et al.}}
\def\papo{Shapelets~I}
\def\papt{Shapelets~II}
\newcommand{\com}[1]{#1}
\def\gs{\mathrel{\raise1.16pt\hbox{$>$}\kern-7.0pt 
\lower3.06pt\hbox{{$\scriptstyle \sim$}}}}         
\def\ls{\mathrel{\raise1.16pt\hbox{$<$}\kern-7.0pt 
\lower3.06pt\hbox{{$\scriptstyle \sim$}}}}         

\ifoldfss  %

  \ifCUPmtlplainloaded \else
    \NewTextAlphabet{textbfit} {cmbxti10} {}
    \NewTextAlphabet{textbfss} {cmssbx10} {}
    \NewMathAlphabet{mathbfit} {cmbxti10} {} 
    \NewMathAlphabet{mathbfss} {cmssbx10} {} 
  \fi
  \ifAMStwofonts
    \ifCUPmtlplainloaded \else
      \NewSymbolFont{upmath} {eurm10}
      \NewSymbolFont{AMSa} {msam10}
      \NewMathSymbol{\upi}     {0}{upmath}{19}
      \NewMathSymbol{\umu}     {0}{upmath}{16}
      \NewMathSymbol{\upartial}{0}{upmath}{40}
      \NewMathSymbol{\leqslant}{3}{AMSa}{36}
      \NewMathSymbol{\geqslant}{3}{AMSa}{3E}

      \let\leq=\leqslant \let\le=\leqslant
       \let\ge=\geqslant
    \fi
  \fi
\fi 

\ifnfssone
  \newmathalphabet{\mathit}
  \addtoversion{normal}{\mathit}{cmr}{m}{it}
  \addtoversion{bold}{\mathit}{cmr}{bx}{it}
  \newmathalphabet{\mathbfit} 
  \addtoversion{normal}{\mathbfit}{cmr}{bx}{it}
  \addtoversion{bold}{\mathbfit}{cmr}{bx}{it}
  \newmathalphabet{\mathbfss} 
  \addtoversion{normal}{\mathbfss}{cmss}{bx}{n}
  \addtoversion{bold}{\mathbfss}{cmss}{bx}{n}
  \ifAMStwofonts
    \ifCUPmtlplainloaded \else
      \UseAMStwoboldmath
      \makeatletter
      \new@mathgroup\upmath@group
      \define@mathgroup\mv@normal\upmath@group{eur}{m}{n}
      \define@mathgroup\mv@bold\upmath@group{eur}{b}{n}
      \edef\UPM{\hexnumber\upmath@group}
      \new@mathgroup\amsa@group
      \define@mathgroup\mv@normal\amsa@group{msa}{m}{n}
      \define@mathgroup\mv@bold\amsa@group{msa}{m}{n}
      \edef\AMSa{\hexnumber\amsa@group}
      \makeatother
      \mathchardef\upi="0\UPM19
      \mathchardef\umu="0\UPM16
      \mathchardef\upartial="0\UPM40
      \mathchardef\leqslant="3\AMSa36
      \mathchardef\geqslant="3\AMSa3E

      \let\leq=\leqslant \let\le=\leqslant
       \let\ge=\geqslant
    \fi
  \fi
\fi 

\ifnfsstwo
  \DeclareMathAlphabet{\mathbfit}{OT1}{cmr}{bx}{it}
  \SetMathAlphabet\mathbfit{bold}{OT1}{cmr}{bx}{it}
  \DeclareMathAlphabet{\mathbfss}{OT1}{cmss}{bx}{n}
  \SetMathAlphabet\mathbfss{bold}{OT1}{cmss}{bx}{n}
  \ifAMStwofonts
    \ifCUPmtlplainloaded \else
      \DeclareSymbolFont{UPM}{U}{eur}{m}{n}
      \SetSymbolFont{UPM}{bold}{U}{eur}{b}{n}
      \DeclareSymbolFont{AMSa}{U}{msa}{m}{n}
      \DeclareMathSymbol{\upi}{0}{UPM}{"19}
      \DeclareMathSymbol{\umu}{0}{UPM}{"16}
      \DeclareMathSymbol{\upartial}{0}{UPM}{"40}
      \DeclareMathSymbol{\leqslant}{3}{AMSa}{"36}
      \DeclareMathSymbol{\geqslant}{3}{AMSa}{"3E}

      \let\leq=\leqslant \let\le=\leqslant
       \let\ge=\geqslant
    \fi
  \fi
\fi 

\ifCUPmtlplainloaded \else
  \ifAMStwofonts \else 
    \def\upi{\pi}
    \def\umu{\mu}
    \def\upartial{\partial}
  \fi
\fi

\title{Polar Shapelets}
\author[R.~Massey \& A.~Refregier]
{Richard~Massey$^{1}$ and Alexandre~Refregier$^{2}$\\
$^1$ California Institute of Technology, 1200 E. California Blvd., Pasadena, California 91125, U.S.A.; {\tt rjm@astro.caltech.edu} \\
$^2$ Service d'Astrophysique, CEA/Saclay, 91191 Gif-sur-Yvette, France; {\tt refregier@cea.fr} \\}

\date{Accepted ---. Received ---; in original form 24 August 2004.}

\pagerange{\pageref{firstpage}--\pageref{lastpage}}
\pubyear{2003}

\begin{document}

\maketitle

\label{firstpage}

\begin{abstract} The shapelets method for image analysis is based upon the
decomposition of localised objects into a series of orthogonal components with
convenient mathematical properties. We extend the ``Cartesian shapelet''
formalism from earlier work, and construct ``polar shapelet'' basis functions
that separate an image into components with explicit rotational symmetries.
These frequently provide a more compact parameterisation, and can be interpreted
in an intuitive way. Image manipulation in shapelet space is simplified by the
concise expressions for linear coordinate transformations; and shape measures
(including object photometry, astrometry and galaxy morphology estimators) take
a naturally elegant form. Particular attention is paid to the analysis of
astronomical survey images, and we test shapelet techniques upon real data from
the {\it Hubble Space Telescope}. We present a practical method to automatically
optimise the quality of an arbitrary shapelet decomposition in the presence of
observational noise, pixellisation and a Point-Spread Function. A central
component of this procedure is the adaptive choice of the shapelet expansion's
scale size and truncation order. A complete software package to perform shapelet
image analysis is made available on the world-wide web.

\end{abstract}


\section{Introduction} \label{intro}

In the shapelets formalism (Refregier 2003; hereafter \papo), individual objects
in an image are decomposed into weighted sums of orthogonal basis functions. The
particular set of basis functions has been chosen to be mathematically
convenient for image manipulation and analysis. In astronomical images, it also
provides a compact representation for the shapes of galaxies of all
morphological types. Refregier \& Bacon (2003; hereafter \papt) showed how these
properties could be used to measure the slight distortions in galaxy shapes due
to weak gravitational lensing. The elegant behaviour of shapelets under Fourier
transform also enabled Chang \& Refregier (2002) to reconstruct images from
interferometric observations. Massey \etal\ (2004) used shapelets to simulate
realistic astronomical images containing galaxies with complex morphologies. A
classification scheme for galaxy morphologies using Principal Component Analysis
of the shapelet basis set was discussed in that paper and applied to the Sloan
Digital Sky Survey by Kelly \& McKay (2004). A method similar to shapelets has
also been independently suggested by Bernstein \& Jarvis (2002; hereafter
BJ02). 

In this paper, we expand upon the earlier work of \papo, \papt\ and BJ02,
developing the formalism of ``polar shapelets'', in which an image is decomposed
into components with explicit rotational symmetries. Whilst the original
Cartesian shapelets remain useful in certain situations, the polar shapelets,
which are separable in $r$ and $\theta$, frequently provide a more elegant and
intuitive form. We find estimators of an object's flux, position and size, that
form naturally from groups of its polar shapelet coefficients. We calculate the
behaviour of a polar shapelet model during linear coordinate transformations. We
also improve the basic shapelet decomposition by incorporating treatments of
pixellisation, observational noise and point-spread functions, and optimising
the overall quality of image reconstruction while maximising data compression.
To test our method upon real data, we extract isolated galaxies from the Hubble
Deep Fields (Williams \etal\ 1996, 1998; hereafter HDFs). These deep,
high-resolution images from the {\it Hubble Space Telescope} (HST) provide
typical examples of distant galaxies' irregular shapes.

A complete {\tt IDL} software package to perform the image
decomposition and shape analyses presented in this paper can be
downloaded from {\tt http://www.astro.caltech.edu/
$\sim$rjm/shapelets/}.

This paper is laid out as follows. In \S\ref{formalism}, we introduce the
Cartesian and polar shapelet basis functions, and their relation to each other.
In \S\ref{optimise}, we investigate the qualitative effects of varying the
shapelet scale size $\beta$ and set quantitative goals for the optimisation of
this choice. In \S\ref{mk_decomp}, we develop practical techniques to cope with 
the effects of pixellisation, seeing and noise in real data. We then demonstrate
various applications of shapelets: in \S\ref{transformations}, we illustrate the
manipulation of images in terms of their changing polar shapelet coefficients
under coordinate transforms. In \S\ref{estimators}, we construct basic shape
estimators for a shapelet model, including flux, centroid and size measures. In
\ref{hosm}, we develop more advanced shape measures that can be used to
quantitatively distinguish galaxies of various morphological types. We conclude
in \S\ref{conc}.

\section{Shapelets formalism} \label{formalism}

\subsection{Cartesian basis functions} \label{cart}

The shapelet image decomposition method was introduced in \papo, and a
related method has been independently suggested by BJ02. The idea is
to express the surface brightness of an object $f(x,y)$ as a linear
sum of orthogonal 2D functions,

\begin{equation}
\label{eqn:decompose_2d}
  f({\mathbf x}) = \sum_{n_1=0}^{\infty}\sum_{n_2=0}^{\infty}
  f_{n_1,n_2}~\phi_{n_1,n_2}({\mathbf x};\beta)~,
\end{equation}

\noindent where the $f_{n_1,n_2}$ are the ``shapelet coefficients'' to
be determined. The dimensionful shapelet basis functions
$\phi_{n_1n_2}$ are

\begin{figure}
\centering
\epsfig{figure=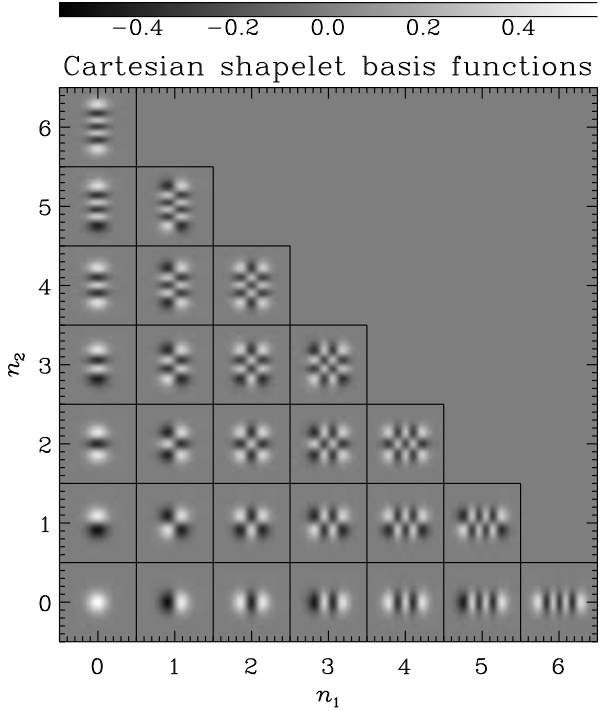,width=84mm} 
\caption{Cartesian
shapelet basis functions, parameterized by two integers $n_1$ and
$n_2$, and here truncated at $n_{\rm max}=6$. An image can be
decomposed into a weighted sum of these functions. This basis is
particularly convenient for many aspects of image analysis and
manipulation commonly used in astronomy and other sciences.}
\label{fig:cart_basis}
\end{figure}

\begin{equation}\label{eqn:bf}
  \phi_{n_1,n_2}({\mathbf x};\beta) \equiv
  \frac
       {
       H_{n_1}\left(\frac{x}{\beta}\right)~
       H_{n_2}\left(\frac{y}{\beta}\right)~
       e^{-\frac{|{\mathbf x}|^2}{2\beta^2}}
      }
      {
       \beta ~ 2^{n} \sqrt{ \pi n_1! n_2!}
      }~,
\end{equation}

\noindent where $H_n({x})$ is a Hermite polynomial of order $n$, and
$\beta$ is the shapelet scale size. These Gauss-Hermite polynomials
form a complete and orthonormal basis set; this ensures that the
shapelet coefficients for any image can be simply and uniquely
determined by evaluating the ``overlap integral''

\begin{equation}
f_{n_1,n_2} = \iint_{\mathbb{R}} f({\mathbf
x})~\phi_{n_1,n_2}({\mathbf x};\beta)~ {\mathrm d}^2x ~.
\label{eqn:lindecompc}
\end{equation}

In practice, a shapelet expansion (eq.~\ref{eqn:decompose_2d}) must be
truncated at a finite order $n_1+n_2 \leq n_{\rm max}$. The array of
shapelet coefficients is sparse for typical galaxy morphologies, which
therefore can be accurately modelled using only a few coefficients. As
shown in \papo, data compression ratios as high as 60:1 can be
achieved for well resolved HST images. Note however, that our choice
of Gauss-Hermite basis functions was not governed by the physics of
galaxy morphology and evolution but by the mathematics of image
manipulation. As we shall see throughout this paper, a shapelet
parameterization is mathematically convenient for many tasks common in
astronomy and other sciences.


\subsection{Polar shapelet basis functions} \label{polar}

\begin{figure} 
\centering
\epsfig{figure=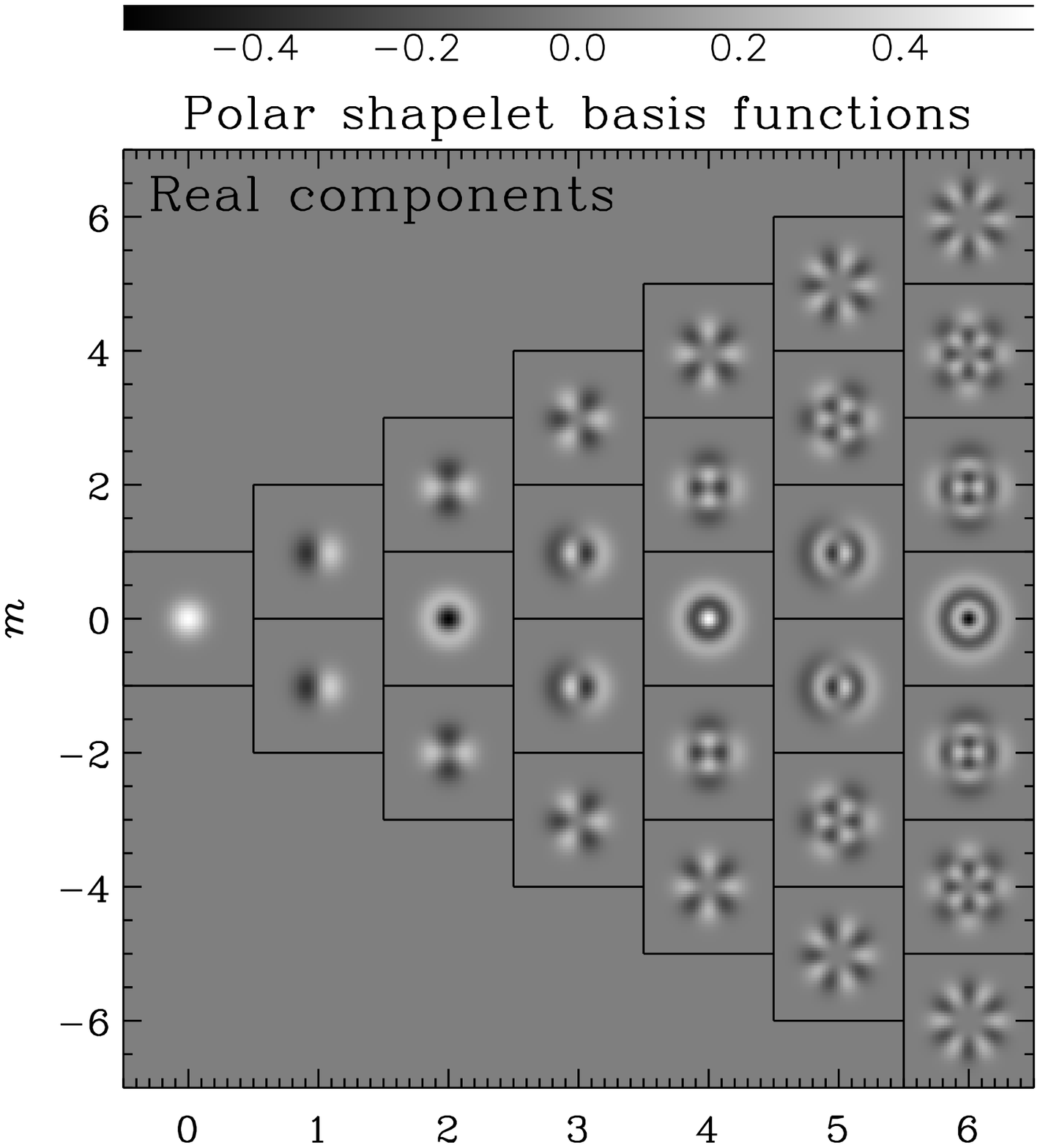,width=84mm} 
\epsfig{figure=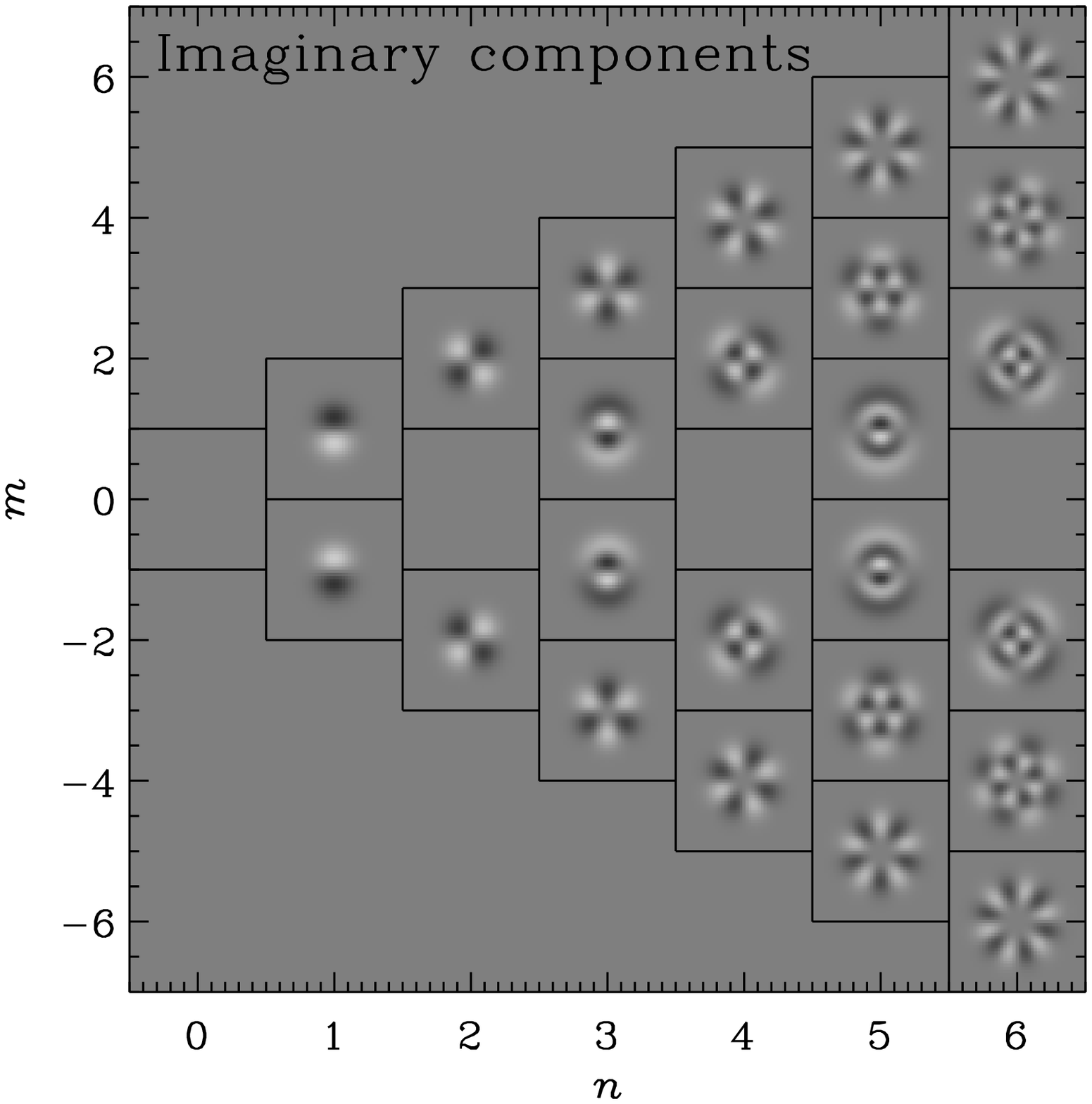,width=84mm}  \caption{Polar
shapelet basis functions. The real components of the complex
functions are shown in the top panel, and the imaginary components in
the bottom. The basis functions with $m=0$ are wholly real. In a
shapelet decomposition, all of the basis functions are weighted by a
complex number, whose magnitude determines the strength of a
component and whose phase sets its orientation.}  \label{fig:basis}
\end{figure}

Polar shapelets were introduced in \papo\ as an orthogonal
transformation of the Cartesian basis states, and were independently
proposed by BJ02. They have all the useful properties of Cartesian
shapelets, and a similar Gaussian weighting function with a given
scale size $\beta$. However, the polar shapelet basis functions are
instead separable in $r$ and $\theta$. This renders many operations
more intuitive, and makes polar shapelet coefficients easy to
interpret in terms of their explicit rotational symmetries.

The polar shapelet basis functions $\chi_{n,m}(r,\theta;\beta)$ are also
parameterized by two integers, $n$ and $m$, and a smooth function
$f(r,\theta)$ in polar coordinates may be decomposed into

\begin{equation} \label{eqn:sseriesp}
f(r,\theta) = \sum_{n=0}^\infty \sum_{m=-n}^{n} f_{n,m}
\chi_{n,m}(r,\theta;\beta) ~.
\end{equation}

\noindent The polar shapelet coefficients $f_{n,m}$ are again given
by the ``overlap integral''

\begin{equation} \label{eqn:lindecompp}
f_{n,m} = \iint_{\mathbb{R}} f(r,\theta)
~\chi_{n,m}(r,\theta;\beta)~r~{\mathrm d}r {\mathrm d}\theta ~.
\end{equation}

BJ02 showed that the ``polar Hermite polynomials'' $H_{n_l,n_r}(x)$,
which were described in \papo, are related to associated Laguerre
polynomials

\begin{equation} \label{eqn:polardef}
L^q_p(x) \equiv \frac{x^{-q}e^x}{p!}~\frac{{\mathrm d}^p}{{\mathrm
d}x^p} \left( x^{p+q}e^{-x} \right) ~,
\end{equation}

\noindent for $n_r>n_l$ by

\begin{equation}
H_{n_{l},n_{r}}(x) \equiv (-1)^{n_{l}}~(n_{l}!)~x^{n_{r}-n_{l}}
L_{n_{l}}^{n_{r}-n_{l}}(x^{2}) ~.
\end{equation}

\noindent Here $n_{l}$ and $n_{r}$ are any non-negative integers. In this paper
we shall instead prefer the simpler $n$, $m$ notation, where $n=n_r+n_l$ and
$m=n_r-n_l$. In this scheme, $n$ can be any non-negative integer, and $m$ can
be any integer between $-n$ and $n$ in steps of two. We truncate the series at
$n\le n_{\rm max}$. Although the only allowed states are those with $n$ and $m$
both even or both odd, this condition will not be written explicitly
alongside every summation for the sake of brevity.

As plotted in figure~\ref{fig:basis}, the dimensionful polar
shapelet basis functions are therefore

\begin{eqnarray}
\chi_{n,m}(r,\theta;\beta) = 
  \frac{(-1)^{\frac{n-|m|}{2}}}{\beta^{|m|+1}}
  \left[\frac{\left(\frac{n-|m|}{2}\right)!}{\pi \left(\frac{n+|m|}{2}\right)! }\right]^{\frac{1}{2}}
  ~\times~~~~~~~~ \nonumber \\
  r^{|m|}
  L_{\frac{n-|m|}{2}}^{|m|} \left(\frac{r^2}{\beta^2}\right)
  e^\frac{-r^{2}}{2\beta^2}
  e^{-im\theta} ~.
\end{eqnarray}

\noindent These are different from the Laguerre expansion used by BJ02 in two
ways. Those are the complex conjugate of our basis functions: {\it i.e.} their
$m$ is equivalent to our $-m$. The Laguerre expansion in BJ02 is also normalised
by one less factor of $\beta$. This dimensionality ensures that, as in the case
of Cartesian shapelets, the polar shapelet basis functions are orthonormal

\begin{equation} \label{eqn:orthop}
\iint_{\mathbb{R}} \chi_{n,m}^*(r,\theta;\beta)~
\chi_{n',m'}(r,\theta;\beta) ~r~{\mathrm d}r {\mathrm d}\theta =
\delta_{n,n'} \delta_{m,m'}
\end{equation}

\noindent as well as complete (see {\it e.g.}\ W\"{u}nsche 1998)

\begin{equation} \label{eqn:polarcomplete}
\sum_{n=0}^\infty \sum_{m=-n}^{n} \chi_{n,m}(r,\theta;\beta)
\chi_{n,m}(r',\theta';\beta)=\delta(r-r')\delta(\theta-\theta')
\end{equation}

\noindent where $\delta$ is the Kronecker delta and the asterisk denotes complex
conjugation. Only those basis functions with $m=0$ contain net flux.

\begin{equation} \label{eqn:intprop}
\iint_{\mathbb{R}} \chi_{n,m}(r,\theta;\beta)
~r~{\mathrm d}r {\mathrm d}\theta = 2 \sqrt{\pi} \beta~
\delta_{m0}~.
\end{equation}

\begin{figure}  \centering
\epsfig{figure=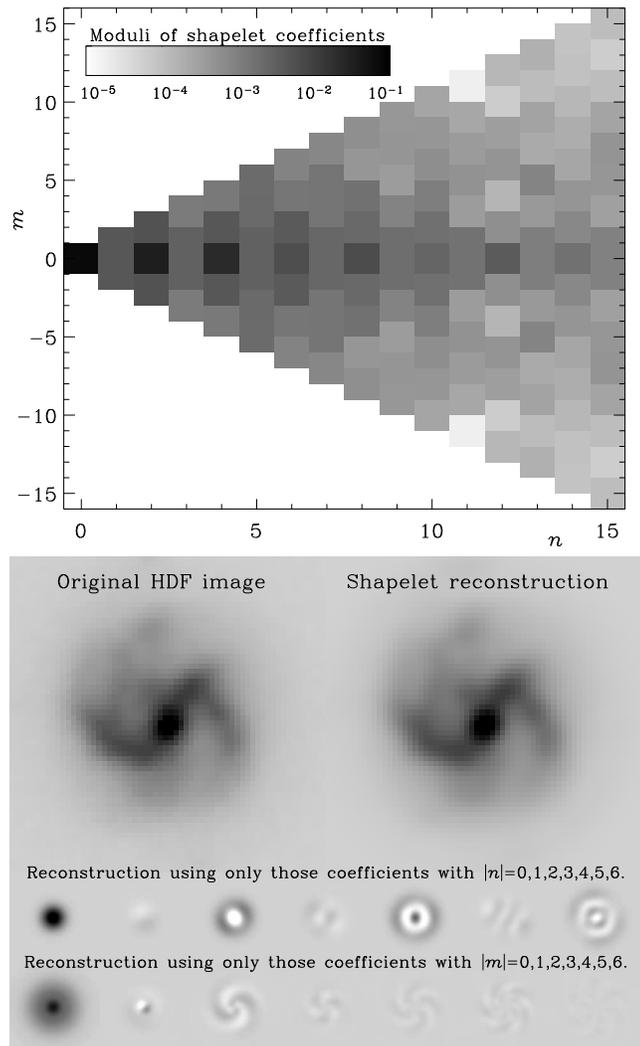,width=84mm} \caption{Example
polar shapelet decomposition of a HDF galaxy. {\it Top panel}: the
moduli of the polar shapelet coefficients, with a logarithmic colour
scale. {\it Bottom panel}: the original galaxy image using a linear
colour scale and its shapelet reconstruction using $n_{\rm max}=20$.
Additional reconstructions are shown using only particular sets of
coefficients, to highlight the contribution of components containing
different symmetries. \label{fig:hstgal}} \end{figure}

Figure~\ref{fig:hstgal} demonstrates the polar shapelet decomposition of a
galaxy found in the HDF. The original image (middle left panel) agrees well with
the reconstruction using $n_{\rm max}=20$. The top panel shows the modulus of
the polar shapelet coefficients as function of the $n$ and $m$ indices. The
dominant coefficients have small values of both indices, demonstrating the
compactness of a polar shapelet representation, and further improved prospects
for data compression. The bottom panel shows the reconstruction of the galaxy
using only coefficients with given values of $n$ or $|m|$, thus highlighting the
contributions of terms with specific rotational symmetries. The off-central
bulge is captured in the $|m|=1$ coefficients, and the main spiral arms in the
$|m|=2$ coefficients. The spiral arms can also be seen as the rotation of the
$n$-only reconstructions with increasing radius. The fainter spiral arms appear
as an interplay of the $|m|=4$ and 5 coefficients.

\subsection{Conversion between Cartesian and polar shapelets}

Cartesian shapelets are real functions, but polar shapelet basis functions
$\chi_{n,m}$ and coefficients $f_{n,m}$ are both complex. However, their
symmetries

\begin{equation}
\label{eqn:chi_sym} \chi_{n,-m}(r,\theta;\beta) =
\chi_{n,m}^{*}(r,\theta;\beta) = \chi_{n,m}(r,-\theta;\beta) ~,
\end{equation}

\noindent simplify matters somewhat if we are concerned only with the
representation of real functions $f({\mathbf x})$, like the surface brightness
of an image. Equations~(\ref{eqn:orthop}) and (\ref{eqn:chi_sym}) imply that
$f({\mathbf x})$ is real if and only if

\begin{equation}
\label{eqn:real} f_{n,-m}=f_{n,m}^{*} ~.
\end{equation}

\noindent Coefficients with $m=0$ are thus wholly real. All polar shapelet
coefficients are paired with their complex conjugate on the other side of the
line $m=0$. Therefore, even though the polar shapelet coefficients $f_{n,m}$ are
generally complex, the number of independent parameters in the shapelet
decomposition of a real function is conserved from the Cartesian case. 

A set of Cartesian shapelet coefficients $f_{n_1,n_2}$ with $n_1+n_2\leq n_{\rm
max}$ can be transformed, into polar shapelet representation with $n\leq n_{\rm
max}$, using
\begin{eqnarray} \label{eqn:cart_polar}
f_{n,m} ~ = ~
 2^{-\frac{n}{2}} i^{m}
 \left[ \frac{n_1 ! n_2 !}{\left(\frac{n+m}{2}\right) ! \left(\frac{n-m}{2}\right) !} \right]^{\frac{1}{2}}
 \delta_{n_1+n_2,n} ~ \times ~~~~~~~~~~~~~~~ \\
\sum_{n_r'=0}^{n_r} \sum_{n_l'=0}^{n_l} i^{m'}
 \left( \begin{array}{c} \left(\frac{n+m}{2}\right) \\ n_{r}' \end{array} \right) \left(
 \begin{array}{c} \left(\frac{n-m}{2}\right) \\ n_{l}' \end{array} \right) 
 \delta_{n_l'+n_r',n_1} ~ f_{n_1,n_2} ~. \nonumber
\end{eqnarray}

\noindent The particular choices of truncation scheme for Cartesian and polar
shapelets now make sense as a way to keep this mapping one-to-one.

\section{Choice of shapelet scale size} \label{optimise}

A shapelet decomposition requires values for the scale size $\beta$ and for the
centre of the basis functions ${\mathbf x}_c$ to be specified in advance.
Choosing the centre is relatively easy: there are many methods well-known in
the astronomical literature to accurately determine astrometry from the
flux-weighted moments of objects. However, the selection of $\beta$ is a less
well-posed problem. In this section, we shall first use some properties of
polar shapelets to describe the effect that the choice of the scale size has
upon a shapelet decomposition. We shall then set quantitative criteria for the
selection of $\beta$ in arbitrary galaxy images that we can implement in a
practical algorithm.

Note that the selection of $\beta$ will be linked to the selection of $n_{\rm
max}$. As shown in \papo\ \S2.4, these two parameters determine the maximum
extent $\theta_{\rm max}$ and finest resolution $\theta_{\rm min}$ that can be
successfully captured by a shapelet model. If $n_{\rm max}\rightarrow\infty$,
any object can be represented using any scale size $\beta$. But if the shapelet
expansion is truncated at finite $n_{\rm max}$, the shape information needs to
be more efficiently contained within fewer coefficients. It is clearly
desirable in this situation to select a scale size $\beta$ that compresses
information, and lets us store the smallest possible number of coefficients.

\subsection{Radial profiles}\label{profile}

Our discussion can be simplified by initially considering only the
radial profile of an object, thus reducing the task to a
one-dimensional problem. Let us consider an object with surface
brightness $f({\bf x})$. The object's radial profile
$\overline{f}(r)$ is its brightness averaged in concentric rings
about its centre, {\it i.e.}

\begin{equation} \label{eqn:fbar}
\overline{f}(r) = \frac{1}{2\pi} \int_0^{2\pi} f(r,\theta) ~~ 
{\mathrm d}\theta ~.
\end{equation}

\noindent With the object decomposed into polar shapelets
as in equation~(\ref{eqn:lindecompp}), it is easy to show that this
is given by

\begin{equation} \label{eqn:fbars}
\overline{f}(r) = \sum_{n}^{\rm even} f_{n0}~\chi_{n0}(r;\beta) ~.
\end{equation}

\noindent This simple expression results from the fact that only the
$m=0$ basis functions are invariant under rotations. These are given
by

\begin{equation} \label{eqn:chino}
\chi_{n0}(r;\beta) = \frac{(-1)^{n/2}}{\beta\sqrt{\pi}}
L_{\frac{n}{2}}^0(r^{2}/\beta^{2}) e^{-r^{2}/2\beta^{2}} ~.
\end{equation}

\noindent The first few rotationally-invariant basis functions are
written explicitly in table~\ref{tab:polarchi}.

\begin{table}
\centering 
\caption{The first few rotationally-invariant polar Shapelet basis functions.}
\begin{minipage}{140mm} 
\hspace{-6mm}\begin{tabular}{rl} 
\hline
$\chi_{0,0}(r)$&\hspace{-4mm}$=\frac{1}{\beta\sqrt{\pi}}~
e^\frac{-r^2}{2\beta^2}$ \\
$\chi_{2,0}(r)$&\hspace{-4mm}$=\frac{-1}{\beta\sqrt{\pi}}
\left[1-\frac{r^2}{\beta^2}\right]
e^\frac{-r^2}{2\beta^2}$ \\
$\chi_{4,0}(r)$&\hspace{-4mm}$=\frac{1}{\beta\sqrt{\pi}}
\left[1-2\frac{r^2}{\beta^2}+
\frac{1}{2}\frac{r^4}{\beta^4}\right]
e^\frac{-r^2}{2\beta^2}$ \\
$\chi_{6,0}(r)$&\hspace{-4mm}$=\frac{-1}{\beta\sqrt{\pi}}
\left[1-3\frac{r^2}{\beta^2}
+\frac{3}{2}\frac{r^4}{\beta^4}
-\frac{1}{6}\frac{r^6}{\beta^6}
\right]
e^\frac{-r^2}{2\beta^2}$ \\
$\chi_{8,0}(r)$&\hspace{-4mm}$=\frac{1}{\beta\sqrt{\pi}}
\left[1-4\frac{r^2}{\beta^2}
+\frac{6}{2}\frac{r^4}{\beta^4}
-\frac{4}{6}\frac{r^6}{\beta^6}
+\frac{1}{24}\frac{r^8}{\beta^8}
\right]
e^\frac{-r^2}{2\beta^2}$ \\
$\chi_{10,0}(r)$&\hspace{-4mm}$=\frac{-1}{\beta\sqrt{\pi}}
\left[1-5\frac{r^2}{\beta^2}
+\frac{10}{2}\frac{r^4}{\beta^4}
-\frac{10}{6}\frac{r^6}{\beta^6}
+\frac{5}{24}\frac{r^8}{\beta^8}
+\frac{1}{120}\frac{r^{10}}{\beta^{10}}
\right]
e^\frac{-r^2}{2\beta^2}$ \\
\hline
\end{tabular}
\end{minipage}
\label{tab:polarchi} 
\end{table}

\begin{figure} 
\centering
\epsfig{figure=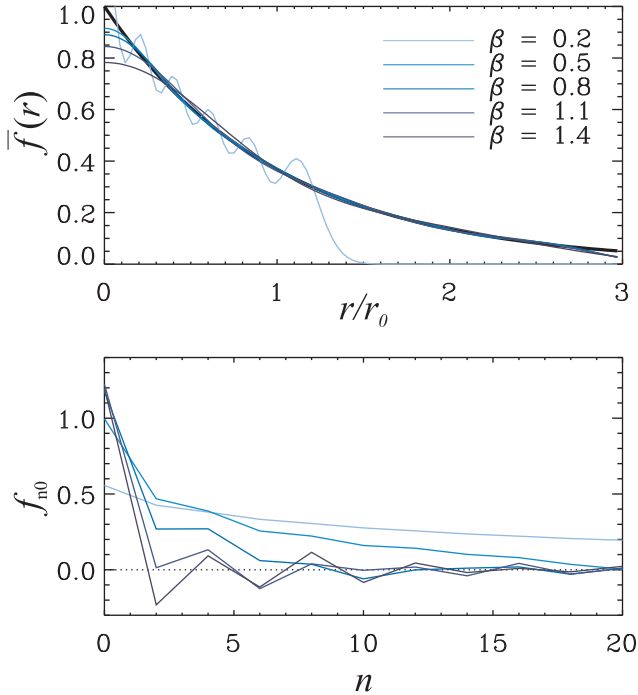,width=84mm}
\caption{Decomposition of an exponential profile into radial polar
shapelets. {\it Top panel:} The thick dark line shows the input
exponential profile. The reconstructed profile is shown for different
values of the shapelet scale $\beta$ with $n_{\rm max}=20$. {\it
Bottom panel}: the corresponding shapelet coefficient profile
$f_{n0}$ versus shapelet order $n$. \label{fig:exprof}} \end{figure}

As a concrete example, we consider the decomposition of galaxy images from the
Hubble Deep Fields (Williams \etal\ 1996, 1998). The mean radial profile of
spiral galaxies is typically exponential, $f(r) \propto e^{-r/r_{0}}$, with some
characteristic radius scale $r_{0}$. Figure~\ref{fig:exprof} shows the shapelet
reconstruction of an exponential radial profile using various values of $\beta$,
with $n_{\rm max}=20$ and the integral in equation~(\ref{eqn:lindecompp})
calculated numerically. 

As can be seen in the top panel of figure~\ref{fig:exprof}, the quality of the
reconstruction depends on the choice of $\beta$. For small values
($\beta \la 0.4r_{0}$) the reconstruction is oscillatory and cuts
off the profile at large radii ($r \ga 1.5r_{0}$). For large
values ($\beta \ga 0.8r_{0}$), the reconstruction fails to
reproduce the cusp at small radii ($r \la 0.4r_{0}$) and exceeds
the true profile at $r\simeq 0.6r_{0}$.
However, for intermediate values ($0.5r_{0}<\beta<1.1r_{0}$), the
reconstruction is good throughout the range $0.1r_{0} \la r \la
2.8r_{0}$. This range can of course be expanded by including more
shapelet coefficients of higher order. As $n_{\rm
max}\rightarrow\infty$, the input model can be recovered with
arbitrary precision using {\it any} value of $\beta$.

The corresponding behaviour in shapelet space is apparent in the
bottom panel of figure~\ref{fig:exprof}. The $f_{n,0}$ coefficients
can be thought as the profile of the galaxy in shapelet space or the
``shapelet profile''. For low values of $\beta$ the shapelet profile
is very flat, showing that the power is distributed almost evenly
throughout all orders. For $\beta=0.5r_0$, the coefficients $a_{\rm
n,m}$ are seen numerically to be $\propto (n+1)^{-2}$. This will be
an important result for the convergence of shape estimators formed
from series of shapelet coefficients in \S\ref{estimators}.
Convergence is fastest at $\beta \approx 0.8r_0$, with $a_{\rm n,m}
\propto (n+1)^{-2.5}$. For higher values of $\beta$, the signs of
$a_{\rm n,m}$ begin to alternate, and the convergence once more falls
below $\propto (n+1)^{-2}$ at $\beta \approx 1.1r_0$.

\begin{figure} 
\centering
\psfig{figure=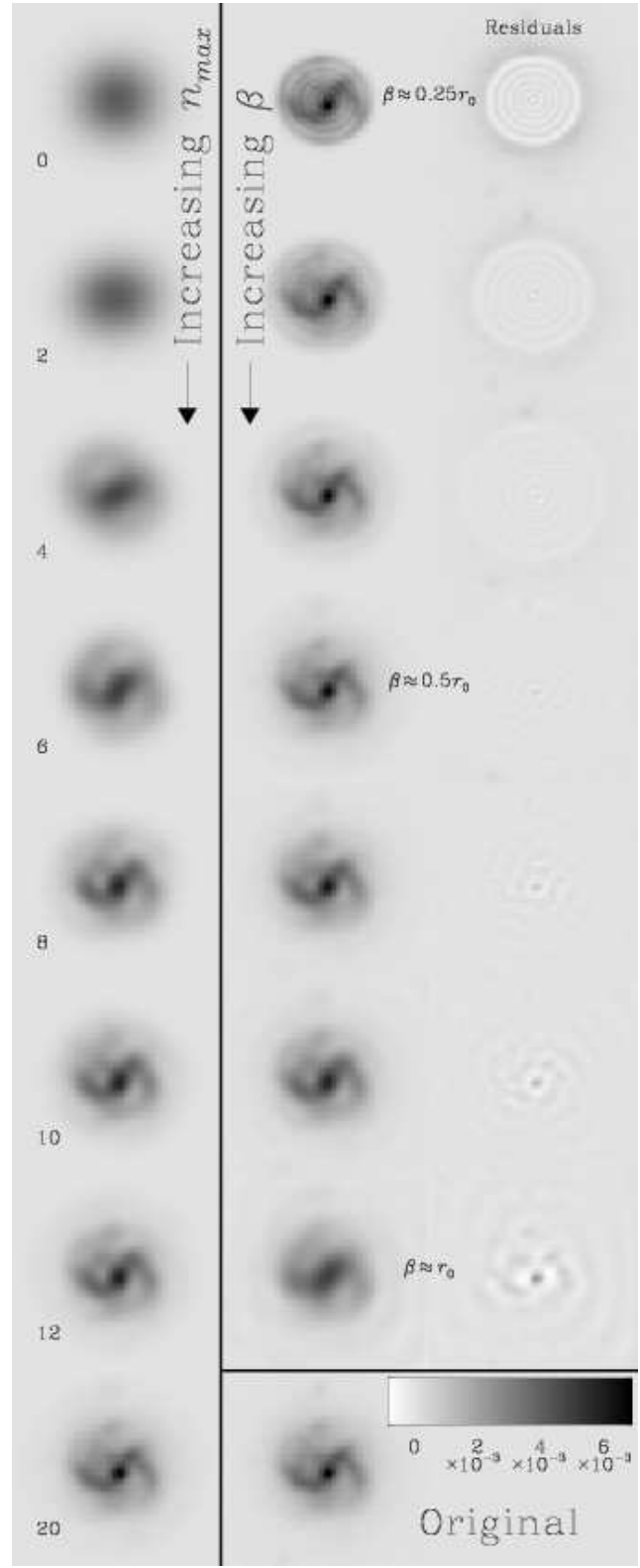,width=84mm}
\caption{Shapelet decomposition of a real spiral galaxy in the HDF.
The best-fit de Vaucouleurs profile has $r_0\simeq12$ pixels. {\it
Left-hand column}: the shapelet model shows growing complexity with
increasing $n_{\rm max}$. For each of these fits, $\beta$ is varied
to minimise the least squares difference between the data and the
model. {\it Right hand columns}: the shapelet decomposition has a
preferred scale size. The residual between the original (in the
bottom-right panel) and these models with fixed $n_{\rm max}=20$ and
varying $\beta$, is smallest with $\beta\simeq0.5r_0$.
\label{fig:varybeta}} \end{figure}

Figure~\ref{fig:varybeta} demonstrates the importance of a proper
choice of the parameters $\beta$ and $n_{\rm max}$ for the practical
decomposition of spiral galaxy in the HDF. Its spiral arms complicate
measurement, but its radial profile is approximately an
exponential with a scale length of $r_0\approx 3\arcsec$ (12~pixels).
The left column shows the growth in complexity of a shapelet model
using increasing $n_{\rm max}$. Note in particular the rotation of
the core ellipticity as $n_{\rm max}$ is increased from 2 to 8 and
higher order moments are included to resolve the spiral arms. In this
column, $\beta$ is allowed to vary in order to minimise the
least-squares difference between the model and the HDF image, shown
at the bottom. The middle column shows shapelet decompositions at
fixed $n_{\rm max}=20$, with varying $\beta$. The residuals are
plotted in the right hand column. As in figure~\ref{fig:exprof}, we
find that the best overall image reconstruction uses
$0.5r_0\la\beta\la0.7r_0$. This is perhaps at the low end of the
range suggested by figure~\ref{fig:exprof} because of the extra
high-frequency detail contained in the spiral arms.

By experimentation we have found a fairly wide range of $\beta$ values that
produce a faithful shapelet reconstruction. The information is then concentrated
into the few lowest shapelet states, with fast convergence to the final model,
and truncation is possible at a computationally acceptable value of $n_{\rm
max}$. We shall now consider ways to formalize this process, and hone our choice
of $x_c$, $\beta$ and $n_{\rm max}$ using quantitative criteria.

\subsection{Existing optimisation methods}
\label{nmaxbeta}

Methods in the literature that face similar choices suggest several distinct
philosophies for the quantitative selection of parameters equivalent to $x_c$,
$\beta$ and $n_{\rm max}$. The suggestions, outlined below, differ both in the
goals set for for an ideal decomposition and the method used to achieve it.

\begin{itemize}

\item \papo\ suggested a geometrical argument using $\theta_{\rm min}$,
$\theta_{\rm max}$: the minimum (PSF or pixel) and maximum (entire
image) sizes on which information exists. This could be iterated
using functional rules on ${\mathbf x}_c$ and $r^2_f$ as defined by
shapelet coefficients. However, the coefficients change as a function
of $n_{\rm max}$, and it is not clear what the rules should be.

\item Van der Marel \& Franx (1993) fit 1D Gauss-Hermite polynomials to
spectral lines. They arbitrarily fix $n_{\rm max}=6$, probably finding this
sufficient because their spectra have relatively high S/N and their lines have
a nearly Gaussian profile. Parameters equivalent to $x_c$ and $\beta$ are
obtained from the best-fitting Gaussian. This also determines $f_{0}$ and in 1D
is equivalent to constraining $f_{1}=f_{2}=0$, {\it i.e.} the derivatives of
the Gaussian with respect to $x_c$ and $\beta$. The number of variables is
reduced and the problem rendered tractable. Unfortunately, this does not help
us in 2D because while both $a_{1\pm 1}$ can be forced to zero by varying
${\mathbf x}_c$, no unique recipe can be found for setting the three $n=2$
states using only one value $\beta$.

\item Van der Marel et al. (1994a,b) relax the constraint on $f_{1}$.
This is an improvement since $f_{1}$ is only the first term of an
expression for the centroid, expanded using {\it all} odd $f_{n}$ in
equation~(\ref{eqn:xc}). Without the higher order corrections, the
true object centroid is moved slightly from the origin: amongst other
things rendering rotations and shear operations more complicated.
Instead, they set $x_c$ from the theoretical rest wavelength of a
line. Unfortunately, astrometric calibration clearly cannot be done
with such accuracy. Nor has the $n=2$ problem been solved.

\item Kaiser, Squires \& Broadhurst (1995) combine fitting with a
stand-alone object detection algorithm, {\tt hfindpeaks}. Translated
into shapelet language, their approach is roughly equivalent to
placing ${\mathbf x}_c$ at data peaks then finding a width $\beta$
such that signal-to-noise ratio $\nu$ in $f_{00}$ is maximised.

\item Bernstein \& Jarvis (2002) propose a similar approach. They prescribe
$\beta$ by requiring $f_{20}=0$, while moving ${\mathbf x}_c$ to ensure $f_{1\pm
1}=0$. Higher coefficients are then determined afterwards by linear
decomposition. To first order, this $\beta$ constraint is equivalent to that for
{\tt hfindpeaks}. This $\beta$ is generally larger than values chosen by our
$\chi^2$ method below, and it can be several times larger for a high
signal-to-noise object containing lots of substructure like the galaxy in
figure~\ref{fig:hstgal}. This method may indeed prescribe the optimal
decomposition for weak lensing as the shear signal in the quadrupole moments
becomes concentrated in one number; however, a predisposition towards particular
states often leads to poor overall image reconstruction and PSF deconvolution,
so it is not necessarily ideal for all applications.

\item Kelly \& McKay (2004) were able to set a fixed {\em physical}
scale of $\beta=$2kpc for galaxies in the Sloan Digital Sky Survey,
where photometric redshifts were available. However, galaxies have a
broad distribution of physical sizes, and it may in fact become more
difficult to interpret a shapelet model derived using this method.

\item Marshall (in preparation) describes a fully Bayesian approach
to applying the shapelet transform in the context of image
reconstruction. Here, ${\mathbf x}_c$, $\beta$ and $n_{\rm max}$ are
varied in order to maximise the evidence (the probability of
observing the data, marginalised over all shapelet coefficients). At
high S/N, this method gives a value of $\beta$  which approaches the
same as that from our $\chi^2$ method below, but otherwise tends to
prefer a fractionally larger $\beta$, conservatively eliminating some
`noise' in favour of a smoother image reconstruction. However, this
is computationally slow, a serious issue when analysing large images.

\end{itemize}

\subsection{Optimisation of image reconstruction}
\label{minchi2}

We shall adopt a choice of $\beta$ and $n_{\mathrm max}$ that is suitable for
many applications, including overall image reconstruction and PSF
deconvolution. Different models will be quantitatively compared via the
overall reconstruction residual

\begin{equation}
\chi^2_r\equiv 
\frac{
\big(f_{\mathrm obs}({\mathbf x})-
     f_{\mathrm rec}({\mathbf x};\beta)\big)^T
   V^{-1}
\big(f_{\mathrm obs}({\mathbf x})-
     f_{\mathrm rec}({\mathbf x};\beta)\big)
}{n_{\rm pixels}-n_{\rm coeffs}}~,
\label{eqn:chi2r}
\end{equation}

\noindent where $f_{\mathrm obs}({\mathbf x})$ is the observed image, and
$f_{\mathrm rec}({\mathbf x};\beta)$ the reconstructed image from the shapelet
model. $V$ is the covariance matrix between pixel values, {\it i.e.} its
diagonal elements are the noise variance in each pixel. We will need to know
this {\it a priori}, or estimate it from blank areas of the image. $n_{\rm
pixels}$ is the number of pixels in the observed image, and $n_{\rm coeffs}$ is
the number of shapelet coefficients used in the model. The residual itself has
variance (Lupton 1993)

\begin{equation}
\sigma^2\big(\chi^2_r\big) = \frac{2}{n_{\rm pixels}-n_{\rm coeffs}} ~.
\end{equation}

\begin{figure}
\centering
\psfig{figure=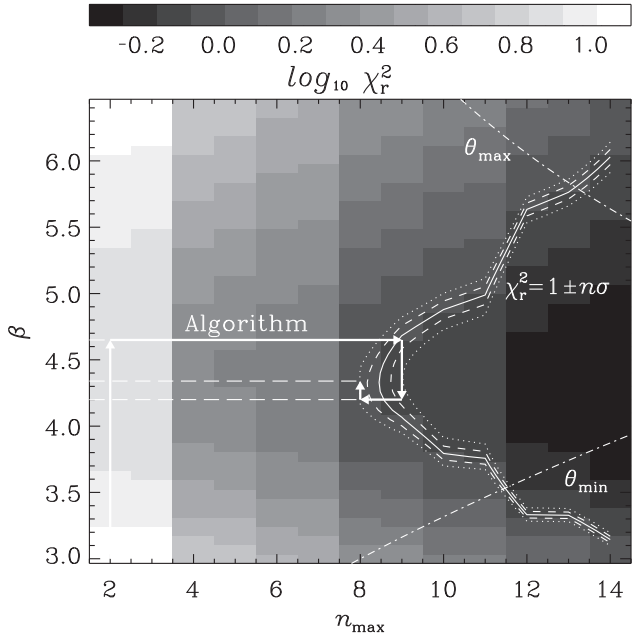,width=84mm}
\caption{ $\chi_r^2$ isocontours on an $n_{\rm max}~vs~\beta$ plane
for an elliptical galaxy in the HDF. The roughly horizontal trough is
typical, with a well-pronounced excursion of the $\chi_r^2$ contours
to lower $n_{\rm max}$ for well-chosen values of $\beta$. The
challenge is to locate the leftmost section of the $\chi_r^2=1$
contour in an automated and efficient way. The arrows show individual
steps (each containing several sub-steps) taken by our optimization
algorithm described in \S\ref{mk_decomp}. Also shown are geometrical
$\theta_{\rm min},\theta_{\rm max}$ constraints and the target
$\chi^2_r=1$ contour.}
\label{fig:betanmax} 
\end{figure}

An example of typical $\chi_r^2$ isocontours on an $n_{\rm max}~vs~\beta$ plane
is shown in figure \ref{fig:betanmax} for an elliptical galaxy from the HDF.
The horizontal trough is present for all galaxies (and many other isolated
objects). This demonstrates that there is indeed an optimum $\beta$ for the
reconstruction of this image. As one might expect, it is roughly independent of
$n_{\rm max}$, but decreases very slightly as more coefficients are added. By
increasing $n_{\rm max}\rightarrow\infty$, the reconstruction can be improved
to arbitrary precision. However, stopping at the $\chi^2_r=1$ contour produces
a model whose residual is consistent with the noise. Additional coefficients
would just model the background noise and should be excluded.

The form of these typical contours thus suggests a unique location in parameter
space. We will choose $\beta$ and $n_{\mathrm max}$ so that the model lies at
the intersection of the trough and the $\chi^2_r=1$ contour, {\it i.e.} at the
left-most point on the contour. To achieve this, we set quantitative goals of

\begin{equation}
  \frac{\partial\chi^2_r}{\partial\beta} = 0 ~,
  \label{eqn:c1}
\end{equation}
\begin{equation}
  {\mathbf x}_c  = 0 ~,
  \label{eqn:c2}
\end{equation}
and
\begin{equation}
  \chi^2_r = 1
  {\mathrm~~or~flattens~out~~}
  \frac{\partial\chi^2_r}{\partial n_{\rm max}}<
        \sigma\left(\chi^2_r\right)
        \approx \sqrt{\frac{2}{n_{pixels}}}~.
  \label{eqn:c3}
\end{equation}

\noindent The first constraint ensures that the scale size is well-suited to
efficiently model the image. The second ensures that the shapelet center matches
the object centroid. The third guarantees that sufficient coefficients ($n_{\rm
max}$) are included to model an object, but with truncation that 'smooths over'
observational noise. A flatness constraint is also included (in the right hand
side of equation~\ref{eqn:c3}). This is particularly important for galaxies with
a near neighbour or for very faint objects that have noisy and fragmented
$\chi_r^2$ contours. In these cases, including additional shapelet coefficients
may not significantly improve a fit, so the series is truncated early. 

We apply extra geometrical constraints to the minimum $\theta_{\rm min}$ and
maximum $\theta_{\rm max}$ scales of the decomposition, to prevent the model
from containing features smaller than the pixel scale or extending off the edge
of an image, where it would be unconstrained. 

\subsection{Automatic optimisation algorithm}
\label{nmaxbetaiteration}

Satisfying the three conditions~(\ref{eqn:c1}), (\ref{eqn:c2}) and
(\ref{eqn:c3}) would ensure that a shapelet decomposition uses the optimum
values of $n_{\rm max}$, $\beta$ and ${\mathbf x}_c$. It is easy to determine
the values of these parameters once the entire $n_{\rm max}~vs~\beta$ plane has
been examined, as in figure~\ref{fig:betanmax}. However, this is a slow
process, so we need a practical algorithm to more efficiently explore this
parameter space, and to iterate rapidly towards our targets. The numerical
implementation of this iteration will inevitably be non-trivial, because it
combines both minimisation and root finding, in a space with one axis discrete.
Here we describe a code that we have developed to repeatedly decompose an
object into shapelets, test the residual, and improve the decomposition
parameters. Its stepwise approach is shown in figure \ref{fig:betanmax}, and
the full code can be downloaded from the world wide web.

\label{catgen}

Objects are first detected in an image using {\tt SExtractor} (Bertin \& Arnouts
1996), a friends-of-friends peak-finding algorithm. After experimenting on
various data sets, we have found the results of {\tt SExtractor} highly
sensitive to input settings. To avoid reliance upon these setting, we use {\tt
SExtractor} as sparingly as possible. We set low detection thresholds in order
to obtain a complete catalogue, and filter out false detections later. We use
the measurement of each object's FWHM to make an initial guess at $\beta$, and
also to set the size of the fixed, circular ``postage stamp'' region that is
extracted around each object. We aim for a postage stamp large enough to contain
the entire object, but small enough to isolate it from its neighbours and to
make the routine computationally efficient. We then use the {\tt SExtractor}
segmentation map to identify pixels in the postage stamp but well away from any
object. These are used to estimate the background noise level, or to locally
renormalise the pixel weight map. Within reasonable limits, the process is
stable with respect to such parameters and we shall not be too concerned as to
the exact {\tt SExtractor} settings. 

Using constant $n_{\rm max}=2$ for speed, $\beta$ is varied in order to
minimise $\chi^2_r$ and satisfy the criterion in equation~(\ref{eqn:c1}), via a
1D version of the Numerical Recipes {\tt AMOEBA} routine: crawling vertically
in figure \ref{fig:betanmax}. During each step of this iteration, the centroid
is simultaneously shifted to re-zero the series in equation~(\ref{eqn:xc}) in
the shapelet coefficients and thus satisfy the criterion in
equation~(\ref{eqn:c2}). Since the calculation of the centroid is independent
of $\beta$ for isolated objects (see \S\ref{hosm}), this part of the iteration
is both stable and fast. Figure \ref{fig:betanmax} also shows the additional
geometrical constraints of $\theta_{\rm min}>0.2$ pixel and $\theta_{\rm max}$
not falling off the edge of the postage stamp. These act as hard boundaries to
the region of parameter space that the amoeba is allowed to explore.

Once the optimum $\beta$ has been found, $n_{\rm max}$ is increased
until the criterion in equation~(\ref{eqn:c3}) is satisfied: crawling
horizontally in figure \ref{fig:betanmax}. The increases are done in
steps of two, because even $n$ states frequently improve the fit more
than odd $n$ states (primarily due to the additiona of a new
$f_{n,0}$ circular state). The value of $n_{\rm max}$ is fine-tuned
to the exact best value at the end. If two values of $n_{\rm max}$
both allow a decomposition with $\chi^2_r=1\pm 1\sigma$, the lower
value is taken.

If the object warranted more coefficients than the initial guess of $n_{\rm
max}=2$, $\beta$ and ${\mathbf x}_c$ are again readjusted at the new $n_{\rm
max}$, using our {\tt 1D AMOEBA} routine. Another $n_{\rm max}$ search then
starts back at $n_{\rm max}=2$ and increases again in steps of two. The
algorithm terminates when either the horizontal or vertical search returns to
the value it started with. All three conditions in equations~(\ref{eqn:c1}),
(\ref{eqn:c2}) and (\ref{eqn:c3}) have then been met. Computation time for each
object increases $\propto n_{\rm max}^4$. On a single 2Ghz processor, our
algorithm takes about 45 minutes to process all of the 3596 objects detected in
the HDF North. 

\begin{figure} \centering
\epsfig{figure=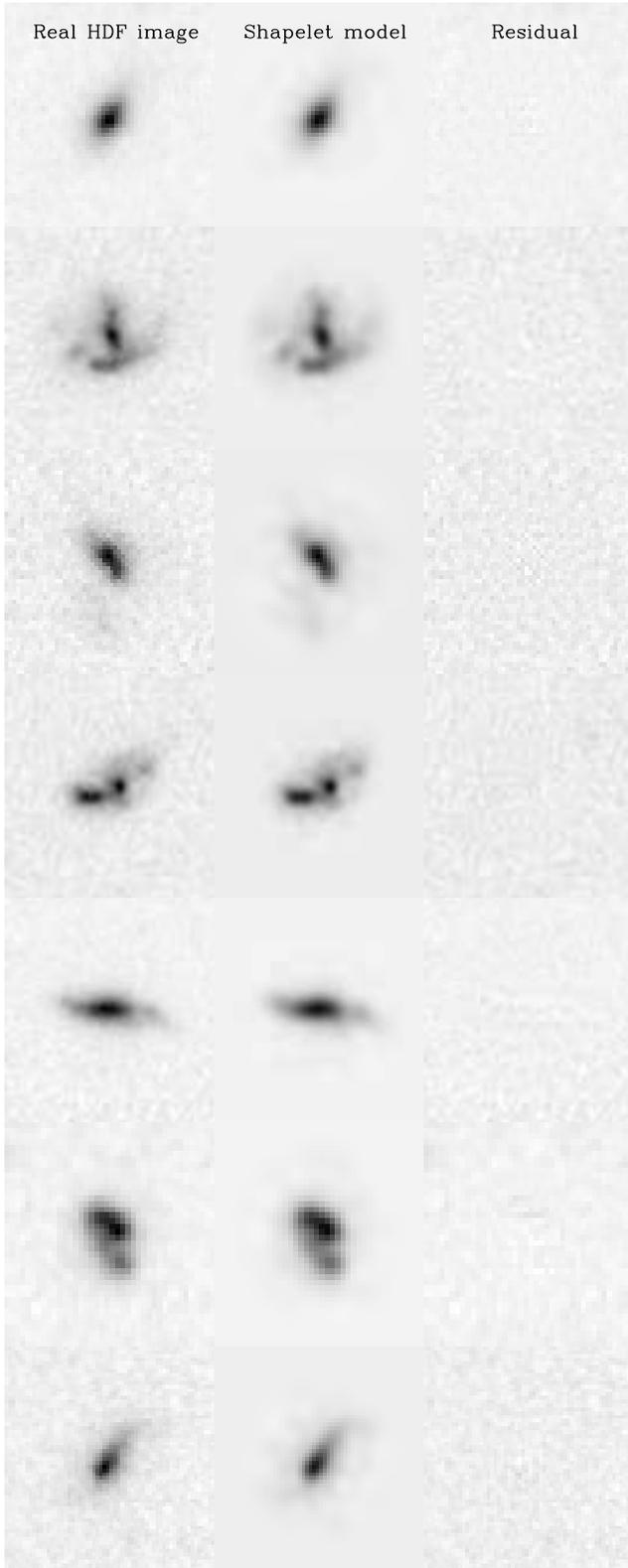,width=84mm} 
\caption{Shapelet models of a selection of HDF galaxies, with their
shapelet scale size $\beta$ and maximum order $n_{\rm max}$
determined automatically. In all cases, the image residuals are
entirely consistent with noise. Our code to perform this task, by
minimising the least squares difference between the model and input
images, is described in the text. \label{fig:decomps}}  \end{figure}

\begin{figure}
\centering
\epsfig{figure=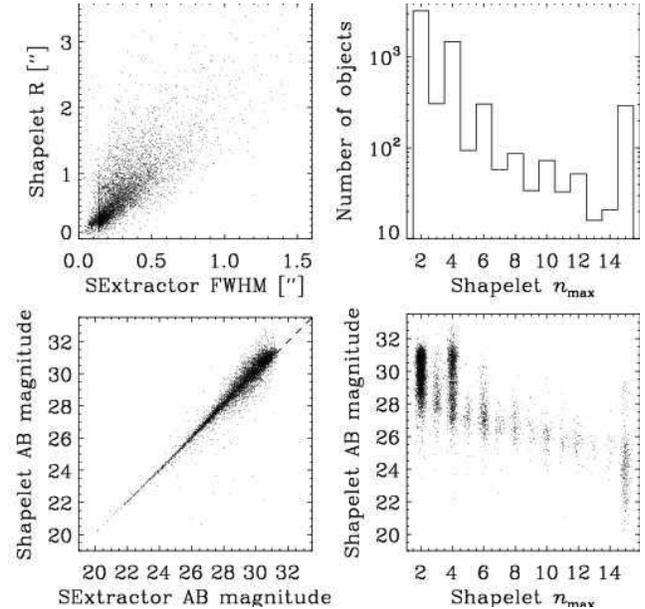,width=84mm}
\caption{The successful recovery of object statistics from the
shapelet parameters of HDF galaxies. For comparison to the {\tt
SExtractor} measurements, shapelet size measurements are shown without
PSF deconvolution. 
In the right-hand panels, galaxies requiring $n_{\rm max}\ge15$ coefficients
have been forced into the final bin, and in the bottom-right panel, points have
been randomly offset a small amount for clarity. \label{fig:mk_obj}} \end{figure}

A selection of reasonably bright HDF galaxies is shown with their shapelet
models in figure~\ref{fig:decomps}. The right-hand column shows the
reconstruction residuals, which are consistent with noise even for irregular
galaxy morphologies. A comparison of their shapelet-based shape estimators to
traditional {\tt SExtractor} measurements is shown in figure \ref{fig:mk_obj}.

\section{Decomposition of real data} \label{mk_decomp}

\subsection{Least squares fitting} \label{lsfit}

Unlike the continuous, analytic formalism presented in \S\ref{formalism}, real
images are complicated by pixellisation, PSF convolution and noise. In order to
incorporate these effects, we shall first adopt a somewhat different approach
to shapelet decomposition than the overlap integrals (\ref{eqn:lindecompc}) and
(\ref{eqn:lindecompp}). We shall instead fit shapelet coefficients to the data
using a least-squares method. Since the model $f_{\rm rec}({\mathbf x})$ in
equation~(\ref{eqn:chi2r}) is linear in the shapelet coefficients, we can solve
for the minimum $\chi^2_r$ solution (\ref{eqn:chi2r}) exactly. We obtain (see
Lupton 1993; Chang \& Refregier 2002)

\begin{equation}
{\mathbf f}_{\mathrm n,m}=(M^TV^{-1}M)^{-1}M^TV^{-1}
  {\mathbf f}_{\mathrm x,y}~, 
\label{eqn:lsfit}
\end{equation}

\noindent where ${\mathbf f}_{\mathrm n,m}$ is a vector of the derived shapelet
coefficients, ${\mathbf f}_{\mathrm x,y}$ the surface brightness in each pixel
arranged as a data vector, $V$ the covariance matrix between pixel values and
$M$ is a matrix of each shapelet basis function evaluated in each pixel. A fit
achieving $\chi^2_r=1$ has successfully modelled all significant spatial
variation in the image, and removed observational noise. 

If the noise per pixel is known, $1\sigma$ confidence limits can be derived on
all of the assigned coefficients using this fitting method (Lupton 1993). If a
complete pixel noise map is available ({\it e.g.}\ from multiple exposures
stacked using {\tt DRIZZLE} software -- Fruchter \& Hook, 2002), it can be used
to down-weight noisy pixels where cosmic rays or hot/cold pixels were present
in some of the exposures. Although the code available on the world-wide web
simply uses a diagonal matrix for $V$ that contains only the noise level in
each pixel, the method is, in general, able to use the full covariance matrix
that contains the amount of covariance between different pixels. In real data,
the flux in adjacent pixels is indeed slightly correlated because of
convolution with the PSF and also because of additional aliasing effects
introduced by {\tt DRIZZLE}. If this effect is important, the pixel-to-pixel
covariances could be estimated from empty regions of an image and included in
the calculation. In particular, this may have a small improvement on statistics
measured from very small objects ({\it c.f.}\ Massey \etal\ 2004).

A constant background level can also be removed using this method, by adding an
undetermined constant to the set of basis functions. Poor flat fielding or
local background gradients near a bright object can also be fit and removed by
adding a plane with variable slope. Although these functions are not strictly
orthogonal, the procedure works well in practice as long as there are
sufficient pixels around the fitted object that contain only background noise.

\subsection{PSF deconvolution} \label{psf}

All real images are inevitably seen after convolution with a Point Spread
Function (PSF). In astronomy, this is typically  caused by atmospheric
turbulence or ``seeing'' (for ground-based observatories), aperture diffraction
at the primary mirror, and imperfect telescope tracking or optics. The
combination of such effects can be measured from the size and shape of stars
observed in an image (because these distant objects would be point-like in the
absence of a PSF), and can be fit with a shapelet model in the same way as the
galaxies. \papo\ presented the matrix operation for convolving an image with a
Gaussian PSF in shapelet space. \papt\ extended this derivation to a general
PSF and demonstrated PSF \emph{de}convolution via matrix inversion. However,
the inversion of the PSF matrix is potentially slow and may be numerically
unstable. Our least-squares fitting method will allow us to elegantly sidestep
this process by convolving the basis functions with the PSF model in advance,
then fitting this new basis set to the data. The returned shapelet model,
reconstructed using the unconvolved basis functions, will be automatically
deconvolved from the PSF.

The formalism for convolution in shapelet space is presented in \papo\ \S4 and
involves three separate scale sizes for three separate objects: $\alpha$ for
the unconvolved model, $\beta$ for the PSF, and $\gamma$ for the convolved
model (there are also corresponding values of $n_{\rm max}^\alpha$, $n_{\rm
max}^\beta$ and $n_{\rm max}^\gamma$). We assume that $\beta$ is known. We can
optimise $\alpha$ as in section \S\ref{minchi2}. However, the choice of
$\gamma$ is a matter entirely internal to the fitting procedure. Just as
before, if $n_{\rm max}^\gamma\rightarrow\infty$, any $\gamma$ will work (but
this time without increasing the number of external free parameters in the
model). In practice, however, it is still necessary to truncate this series
somewhere. Note that $\gamma^2=\alpha^2+\beta^2$ was incorrectly suggested as a
``natural choice'' for this parameter in \papo. Another choice would be
$\gamma=\alpha$, which, with $n_{\rm max}^\gamma=n_{\rm max}^\alpha$, makes the
convolution matrix $P_{\rm n,m}$ symmetric and thus simplifies its calculation.

The optimum values for $\gamma$ and $n_{\rm max}^\gamma$ are in fact obtained
from an argument concerning the information present in shapelet coefficients. A
shapelet model contains information only between a minimum and maximum scales

\begin{equation}
\theta_{\rm min}=\frac{\beta}{\sqrt{n_{\rm max}+1}} ~~~~~~{\mathrm and}~~~~~~
\theta_{\rm max}=\beta\sqrt{n_{\rm max}+1}~.
\end{equation}

\noindent During convolution, $\theta_{\rm min}^\alpha$ and $\theta_{\rm
min}^\beta$ add in quadrature to produce $\theta_{\rm min}^\gamma$; $\theta_{\rm
max}^\alpha$ and $\theta_{\rm max}^\beta$ add similarly to produce $\theta_{\rm
max}^\gamma$; the range of scales on which information is available decreases or
remains constant. Values of $\gamma$ and $n_{\rm max}^\gamma$ can be chosen to
most efficiently capture the information contained between the new scales.
Writing $(n_{\rm max}^\alpha+1)$ as $N_{\alpha}$ etc.\ for brevity, we find 

\begin{equation}
\gamma = \sqrt[4]{ \frac{\big(\alpha^2 N_\alpha + \beta^2 N_\beta\big)
                  \big(\alpha^2 N_\beta  + \beta^2 N_\alpha\big)}
                  {N_\alpha N_\beta} }
\end{equation}

\noindent and

\begin{equation}
n_{\rm max}^\gamma = \sqrt{\frac{\alpha^2 N_\alpha + \beta^2 N_\beta}
                          {\alpha^2 N_\beta + \beta^2 N_\alpha}
                          N_\alpha N_\beta} ~ - ~ 1 ~.
\end{equation}

\begin{figure}
\centering
\epsfig{figure=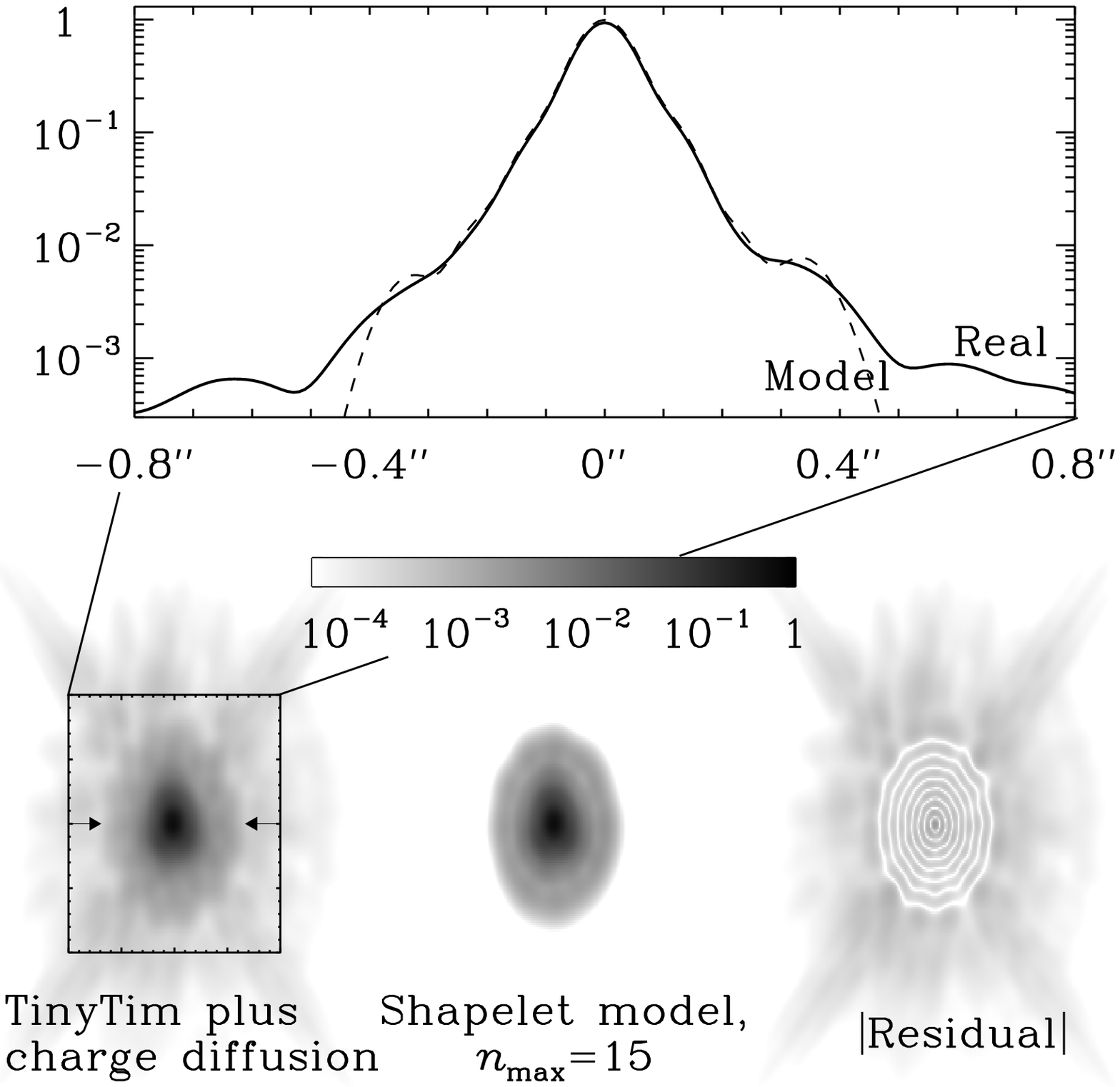,width=84mm}
\epsfig{figure=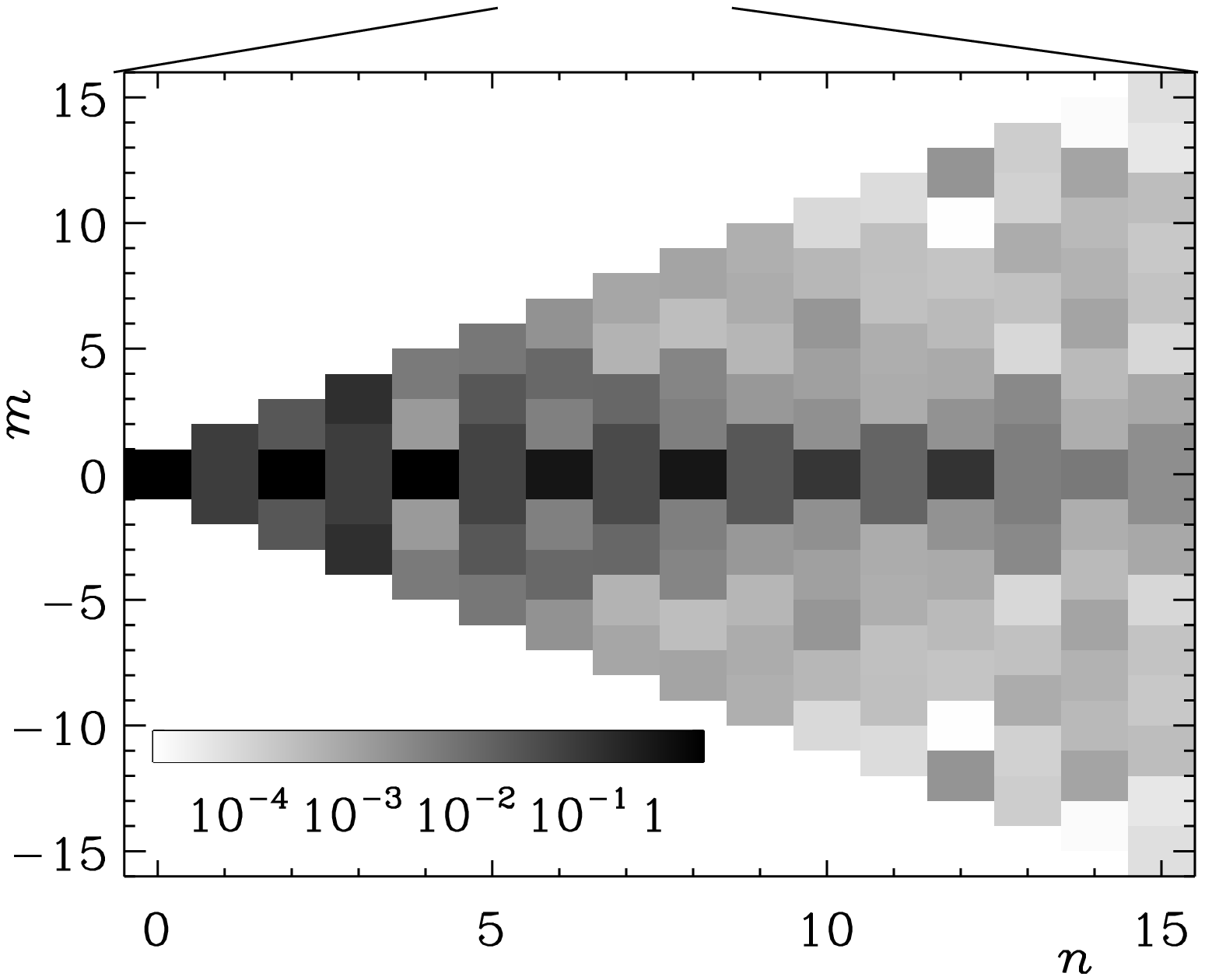,width=84mm} \caption{ Shapelet
model of the {\tt TinyTim} (Krist 1995) WFPC2 PSF plus
charge diffusion. {\it Top panel}: a horizontal slice through
the centre of the PSF. {\it Bottom panel}: the moduli of its polar
shapelet coefficients to $n_{\rm max}=15$. Note that the amplitude
scales are all logarithmic: the core is actually modelled very
successfully out to the second diffraction ring. For
speed we do not bother capturing the wings. \label{fig:psf}}
\end{figure}

The PSFs of cameras on board the {\it Hubble Space Telescope} are well known
and stable. Figure \ref{fig:psf} shows an oversampled {\tt TinyTim} (Krist
1997) model of the {\it Wide-Field Planetary Camera 2} (WFPC2) PSF, raytraced
through an engineering model, plus charge diffusion to simulate photon capture
within the CCD cameras. This is easy to model with shapelets, except for the
fact that its steep cusp and extended wings are intrinsically ill-matched to
the Gaussian around which shapelets are constructed. The PSF is shown in the
figure beside a shapelet decomposition up to $n_{\rm max}^{{\mathrm PSF}}=15$.
This is sufficient to accurately capture the core and the first two diffraction
rings, which are already more than two orders of magnitude below the maximum,
but does not extend to the four faint diffraction spikes or far into the
low-level wings (note that the colour scales are logarithmic). In principle,
this could be further extended at a cost to processor time by using more
shapelet coefficients.

\begin{figure}
\centering
\epsfig{figure=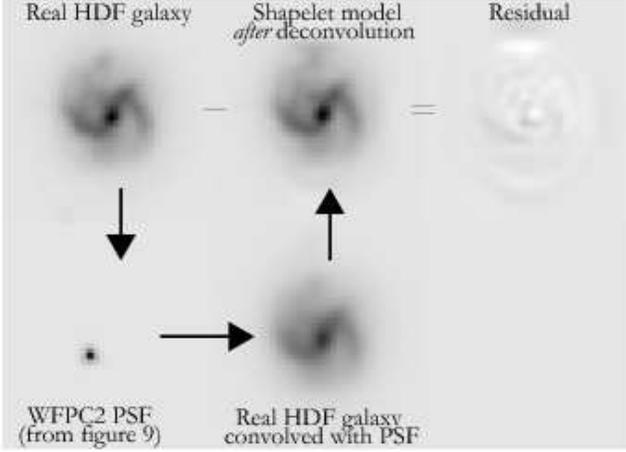,width=84mm}
\caption{Demonstration of deconvolution from an observational PSF.
{\it Top-left panel}: a real HDF galaxy. {\it Bottom-left panel}: the
WFPC2 PSF model, from figure~\ref{fig:psf} but displayed here with a
linear colour scale. {\it Bottom-middle panel}: the galaxy image
convolved with the PSF. This convolution has been performed in real
space, to disassociate the operation from anything involving
shapelets. {\it Top-middle panel}: a shapelet reconstruction and
deconvolution of the galaxy to $n_{\rm max}=20$, obtained from a fit
to the convolved image, assuming knowledge of the PSF. {\it Top-right
panel}: the difference between the true galaxy image and the shapelet
model after deconvolution. This small residual demonstrates the
success of PSF deconvolution using shapelets.}
\label{fig:deconv}
\end{figure}

Figure~\ref{fig:deconv} demonstrates successful PSF deconvolution. A galaxy
from the HDF is convolved with the WFPC2 PSF (in real space). This is treated
as the observed image, and deconvolved from the PSF using a shapelet fit. The
resulting reconstruction is in good agreement with the original galaxy image,
as can be seen from its small residual. Note that the optimum scale size
$\beta$ for the model is slightly lower when PSF deconvolution is performed.
This reflects the need to capture finer details.

\subsection{Pixellisation}\label{phin_int}

Real image data is typically stored in discrete pixels. To link this to the
analytic shapelet formalism, one must either smooth the data or pixellate the
shapelet basis functions. Smoothing the data requires an arbitrary interpolation
scheme to be defined, and resampling the data onto smaller pixels can be very
slow. A better approach is to leave the data alone, and discretize the smooth
shapelet basis functions. This reduces the integrals in equations
(\ref{eqn:lindecompc}) and (\ref{eqn:lindecompp}) to sums over pixel values,
which are fast to compute. However, they are no longer analytically exact. We
therefore need to define a discretization scheme that keeps the basis functions
as orthogonal as possible, and the integrals as accurate as possible.

As pointed out by Berry, Hobson \& Withington (2004), one cannot simply adopt
the value of basis functions at the centre of each pixel. Basis functions that
contain oscillations on scales smaller than the pixel size are sampled in an
essentially random manner. Their discrete versions are then neither
representative of the analytic function nor orthogonal. Degeneracies are
introduced between shapelet coefficients during the decomposition that
unstabilise the inversion of coefficient matrices in the reconstructed model,
and bias quantities like an object's flux. Fortunately, this is rarely a problem
in practical cases, because we can choose $n_{\rm max}$ and $\beta$ in advance
to isolate only those basis functions that contain oscillations on scales larger
than the pixel (or seeing) size. Under these conditions, Berry, Hobson \&
Withington (2004) show that the shapelet basis functions are indeed orthogonal.

We suggest an even safer alternative here. The Cartesian basis functions are
separable in $x$ and $y$, and may be analytically integrated within rectangular
pixels. This is exactly the same process undergone by photons arriving at a CCD,
where the smooth function of a real scene gets binned into digital squares. Once
we have convolved the basis functions with the PSF, and integrated them within
pixels, they can be suitably matched to the data.

To integrate the 2D Cartesian basis functions, first consider the 1D
basis functions from \papo,

\begin{equation}
\phi_{n}(x) \equiv
  \left[ 2^{n}  \pi^{\frac{1}{2}} n! \beta \right]^{-\frac{1}{2}}
  H_{n}\left(\frac{x}{\beta}\right)~e^{-\frac{x^2}{2\beta^2}}~.
\end{equation}

\noindent Integrating by parts and using two well-known identities (see {\it
e.g.} Boas Ch. 12)
\begin{equation}
  H_n(x) = 2xH_{n-1}(x) - 2(n-1)H_{n-2}(x)
\end{equation}
and
\begin{equation}
  \frac{{\mathrm d} H_{n-1}(x)}{{\mathrm d} x} = 2(n-1)H_{n-2}(x)  ~,
\end{equation}

\noindent one can obtain the recurrence relation
\begin{eqnarray}
\label{eqn:intinpixels1}
I_n & \equiv & \int^b_a \phi_{n}(x)~{\mathrm d} x \\
     & = & - \beta \sqrt{\frac{2}{n}}
         \left[ \frac{}{} \phi_{n-1}\left(x\right)\right]^b_a
        + \sqrt{\frac{n-1}{n}} ~ I_{n-2} ~.
\end{eqnarray}

\noindent Finally, note that
\begin{eqnarray}
I_0 & = & \sqrt{\frac{\beta\pi^{\frac{1}{2}}}{2}}
          \left[ \frac{}{} {\mathrm erf}\left(\frac{x}{\beta\sqrt{2}}\right) \right]^b_a ~{\mathrm and} \\
I_1 & = & - \beta\sqrt{2}\left[ \frac{}{} \phi_{0}\left(x\right)\right]^b_a ~.
\end{eqnarray}

\noindent This supplies all the necessary integrals. Since the 2D Cartesian
basis functions are separable in $x$ and $y$, it is easy to extend this
derivation to integrate within square CCD pixels:

\begin{equation}
\label{eqn:intinpixels}
I_{n_1,n_2} = \int^{b_1}_{a_1}\int^{b_2}_{a_2}\phi_{n_1}(x)\phi_{n_2}(y)
                ~{\mathrm d} x ~{\mathrm d} y
            = I_{n_1} \times I_{n_2}
\end{equation}

\noindent where, if there is no `dead zone' around the edge of a
pixel, $(b_1-a_1)\times(b_2-a_2)$ is the angular size of a pixel.  A
missing pixel border, due for instance to electronics which is
unresponsive to light, can be included by altering the limits on the
integral.

We can either use this result to obtain a model in Cartesian shapelet space,
which can later be converted to a polar shapelet representation using
equation~(\ref{eqn:cart_polar}), or we can integrate the polar shapelet basis
functions within pixels using the same equations. This integration is a
particularly important advance for small galaxies or for shapelet basis
functions at high-$n$, that can contain oscillations smaller than a single
pixel.

The symmetries of polar shapelets can also be used to integrate models within
circular apertures using equations~(\ref{eqn:cs1}) to
(\ref{eqn:int0rchinclosed}).


\section{Coordinate transformation} \label{transformations}

Image manipulation via linear transformations is simple in shapelet
space. As in \papo, let us consider an infinitesimal coordinate
transformation ${\mathbf x} \rightarrow (1+{\mathbf \Psi}) {\mathbf
x} + {\mathbf \epsilon}$, where ${\mathbf
\epsilon}=\{\epsilon_{1},\epsilon_{2}\}$ is a displacement and
${\mathbf \Psi}$ is a $2\times2$ matrix parametrized as

\begin{equation} \label{eqn:psi_params}
{\mathbf \Psi} = \left( \begin{array}{cc} \kappa +\gamma_{1} &
\gamma_{2} - \rho \\ \gamma_{2} + \rho & \kappa - \gamma_{1} \\
\end{array} \right).
\end{equation}

\noindent The parameters $\rho$, $\kappa$, $\epsilon$ and
$\gamma_{i}$ correspond to infinitesimal rotations, dilations,
translations and shears.

An image transforms as
$f({\mathbf x}) \rightarrow f'({\mathbf x}) \simeq f({\mathbf
x}-{\mathbf \Psi}{\mathbf x}-{\mathbf \epsilon})$, which can
be written as

\begin{equation}
f' \simeq (1 + \rho \hat{R}+ \kappa \hat{K} + \gamma_{j} \hat{S}_{j}
+ \epsilon_{i} \hat{T}_{i}) f,
\label{eqn:smalltrans}
\end{equation}

\noindent where $\hat{R}$, $\hat{K}$, $\hat{S}_{i}$ and $\hat{T}_{i}$ are the
operators generating rotation, convergence, shears and translations,
respectively. We adopt a notation from weak gravitational lensing, where a
``convergence'' $\kappa$ corresponds to a change in an object's radius by a
factor $(1-\kappa)^{-1}$. These transformations can be viewed as a mapping of
$f_{n,m}$ coefficients in shapelet space. For example, a finite rotation is

\begin{equation} \label{eqn:oprotate}
\hat{R} : f_{n,m} \rightarrow f_{n,m}' = f_{n,m}~e^{im\rho}~,
\end{equation}

\noindent so a rotation through $180^\circ$ can be written as

\begin{equation} \label{eqn:oprot180}
\hat{R}_{180^\circ} : f_{n,m} \rightarrow f_{n,m}' = (-1)^m~f_{n,m}~.
\end{equation}

An (infinitesimal) dilation can be performed in polar shapelet space
by mapping the shapelet coefficients as

\begin{eqnarray} \label{eqn:opdilatesb}
\hat{K}:f_{n,m} \rightarrow f_{n,m}' & = & (1+\kappa)~f_{n,m} \\
 & + & \frac{\kappa}{2}\sqrt{(n-m)(n+m)}~f_{n-2,m} \nonumber \\
 & - & \frac{\kappa}{2}\sqrt{(n-m+2)(n+m+2)}~f_{n+2,m}~. \nonumber
\end{eqnarray}

\noindent The shapelet model may require more coefficients after this
transformation. Note that this dilation operation increases both the flux
and the image area by a factor $1+2\kappa$, thus conserving surface
brightness. To instead perform a dilation that conserves the total
{\em flux}, divide the right hand side of
equation~(\ref{eqn:opdilatesb}) by this factor. To first order, this
is 

\begin{eqnarray} \label{eqn:opdilateflux}
\hat{K}:f_{n,m} \rightarrow f_{n,m}' & = & (1-\kappa)~f_{n,m} \\
 & + & \frac{\kappa}{2}\sqrt{(n-m)(n+m)}~f_{n-2,m} \nonumber \\
 & - & \frac{\kappa}{2}\sqrt{(n-m+2)(n+m+2)}~f_{n+2,m}~. \nonumber
\end{eqnarray}

\noindent In \S\ref{estimators}, we shall ensure that shape estimators
for a shapelet model are independent of the scale factor chosen for the
decomposition by ensuring that the estimators are unchanged under this mapping.

Rather than these first-order approximations, finite dilations can be performed
to all orders using the rescaling matrix in the appendix of \papo. This is
identical to the convolution matrix, but the image is convolved with a
$\delta$-function.

Shears and translations can be performed using
\begin{eqnarray} \label{eqn:opshear}
\hat{S}:f_{n,m} \rightarrow f_{n,m}' = f_{n,m} ~~~~~~~~~~~~~~~~~~~~~~~~~~~~~~~~~~~~~~~~~~~~~~~~~~~~~~ \\
+ \frac{\gamma_1+i\gamma_2}{4}\left\{ \sqrt{(n+m)(n+m-2)}~f_{n-2,m-2} \right. ~~~~ \nonumber \\
- \left. \sqrt{(n-m+2)(n-m+4)}~f_{n+2,m-2} \right\} \nonumber \\
+ \frac{\gamma_1-i\gamma_2}{4}\left\{ \sqrt{(n-m)(n-m-2)}~f_{n-2,m+2} \right. ~~~~ \nonumber \\
- \left. \sqrt{(n+m+2)(n+m+4)}~f_{n+2,m+2} \right\} \nonumber
\end{eqnarray}

\noindent and
\begin{eqnarray} \label{eqn:optranslate}
\hat{T}:f_{n,m} \rightarrow f_{n,m}' = f_{n,m} ~~~~~~~~~~~~~~~~~~~~~~~~~~~~~~~~~~~~~~~~~~~~~~~~~~~~~~~ \\
+ \frac{\epsilon_1+i\epsilon_2}{2\sqrt{2}}\left\{ \sqrt{(n+m)}~f_{n-1,m-1} \right. ~~~~~ \nonumber \\
- \left. \sqrt{(n-m+2)}~f_{n+1,m-1} \right\} ~~ \nonumber \\
+ \frac{\epsilon_1-i\epsilon_2}{2\sqrt{2}}\left\{ \sqrt{(n-m)}~f_{n-1,m+1} \right. ~~~~~ \nonumber \\
- \left. \sqrt{(n+m+2)}~f_{n+1,m+1} \right\} ~, \nonumber
\end{eqnarray}

\noindent with the translation specified in units of $\beta$.

Other image manipulations can also be represented as mappings of
shapelet coefficients. Changes of flux by a factor $B$ are
trivially implemented by the mapping

\begin{equation} \label{eqn:opflux}
\hat{B}:f_{n,m} \rightarrow f_{n,m}' = B \times f_{n,m}~.
\end{equation}

\noindent It is also possible to circularise an object with the
mapping (see \S\ref{profile})

\begin{equation} \label{eqn:opcirc}
\hat{C}:f_{n,m} \rightarrow f_{n,m}' = f_{n,m} ~ \delta_{m0} ~,
\end{equation}


\noindent or to flip an object's parity by reflection in the $x$-axis
using

\begin{equation} \label{eqn:opparity}
\hat{P}:f_{n,m}\rightarrow f_{n,m}'=f_{n,m}^*~.
\end{equation}

\noindent Combining this $\hat{P}$ with the rotation operator
allows reflections to be performed in any axis. 

\begin{figure} 
\centering
\epsfig{figure=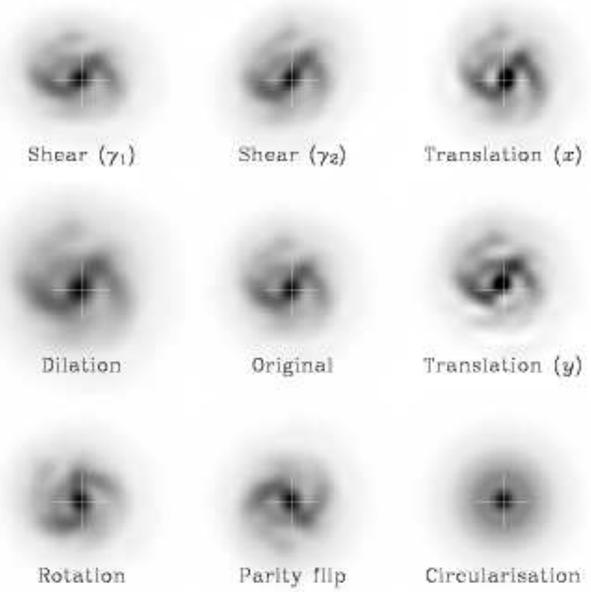,width=84mm}
\caption{Some simple operations applied to a real galaxy image, by
using the polar shapelet ladder operators or coefficient mappings as
described in the text. The central image is the original galaxy.
Starting at the bottom-left and proceeding clockwise, the other
images show rotation by $40^\circ$, dilation of $\kappa=0.15$,
shears of $\gamma=10\%$, translations, circularisation and reflection
in the $x$-axis.} 
\label{fig:ops} 
\end{figure}

These operators' actions are demonstrated upon a real galaxy image in figure
\ref{fig:ops}.

\section{Object shape measurement} \label{estimators}

The above symmetries of the polar shapelet basis functions can be used to
identify combinations of shapelet coefficients that measure an object's flux
(photometry), centroid position (astrometry) and size. Similar weighted
combinations of Cartesian shapelet coefficients were found in \papo, but we find
the interpretation of polar shapelets more intuitive, and the expressions below
are usually more simple than their Cartesian equivalents. For example, the
rotationally invariant part of an object is isolated into its $m=0$
coefficients. The linear offset of an object from the origin is described by its
$m=\pm1$ coefficients and the ellipticity of an object by its $m=\pm2$
coefficients. In the latter cases, the magnitude of the coefficients indicate an
amplitude, and the phases a direction.

\subsection{Photometry} \label{photometry}

Practical measurements of objects' flux usually introduce a Gaussian or top-hat
weight function in order to limit contamination from surrounding noise and
nearby objects. The flux of a shapelet model inside a circular aperture can be
calculated using only the coefficients with $m=0$. All other coefficients
correspond to basis functions with positive and negative regions that cancel out
under integration around $\theta$. From equations~(\ref{eqn:fbars}) and
(\ref{eqn:chino}) for an object's radial profile, we find that

\begin{eqnarray} \label{eqn:cs1}
  \int_0^{2\pi}\int_0^Rf(r)~r~{\mathrm d}r{\mathrm d}\theta
= (4\pi)^{\frac{1}{2}} \beta~\sum_n^{\mathrm even}f_{n,0}~I_n ~,
\end{eqnarray}

\noindent where

\begin{eqnarray} \label{eqn:cs2}
  I_n=\frac{(-1)^{\frac{n}{2}}}{\beta^{2}}\int_0^R
      L_{\frac{n}{2}}^0\left(\frac{r^2}{\beta^2}\right)
      e^\frac{-r^2}{2\beta^2}~r~{\mathrm d}r ~.
\end{eqnarray}

\noindent Using relation~(\ref{eqn:laguerrerecursion2}) to integrate by this
parts, we can find a recursion relation\footnote{Thanks to Mark Coffey for help
deriving this expression.}

\begin{equation} \label{eqn:int0rchin}
  I_n=(-1)^{\frac{n}{2}}\left\{
      1-L_{\frac{n}{2}}^0\left(\frac{R^2}{\beta^2}\right)e^\frac{-R^2}{2\beta^2}
      -2\sum_{i=0}^{\frac{n-2}{2}}(-1)^iI_{2i}
      \right\} ~,
\end{equation}

\noindent and a closed form

\begin{eqnarray} \label{eqn:int0rchinclosed}
  I_n = 1-e^\frac{-R^2}{2\beta^2}\Bigg\{
            2\sum_{i=0}^{n/2}~(-1)^i~L^0_i\left(\frac{R^2}{\beta^2}\right) ~~~~~~~~~~~~~~~~~~~~~~~~ \nonumber \\
            -(-1)^{\frac{n}{2}}~L^0_{\frac{n}{2}}\left(\frac{R^2}{\beta^2}\right)
	    \Bigg\} ~.
\end{eqnarray}

However, the imposition of a integration boundary is unnecessary with shapelets
because the model is analytic and noise-free. In the limit of
$R\rightarrow\infty$, we obtain a simple expression for the total flux in a
shapelet model

\begin{equation}
\label{eqn:f}
F \equiv \iint_{\mathbb{R}} f({\mathbf x})~{\mathrm d}^{2}x ~ = ~
(4\pi)^{\frac{1}{2}} \beta ~ \sum_{n}^{\rm even} f_{n0}~,
\end{equation}

\noindent a result that can also be recovered by transforming the sum over
Cartesian shapelet coefficients from \papo\ into polar shapelet space via
equation~(\ref{eqn:cart_polar}). Cartesian shapelet models can also be
integrated within square apertures using equations~(\ref{eqn:intinpixels1}) to
(\ref{eqn:intinpixels}).

This extrapolation to large radii does rely upon the faithful representation of
an object by a shapelet expansion, and the removal of its noise via series
truncation. Such truncation restricts the basis functions' completeness, and a
weight function (constructed from a combination of the allowed basis functions)
akin to a ``prior probability'' is subtly implicit inside our fitting procedure.
However, a fitting method like ours can beat a direct, pixel-by-pixel
measurement. Our fit is able to include flux from the extended wings of an
object, by integrating it over a large area, even when the signal lies beneath
the noise level in any individual pixel. The wings of galaxies in
figure~\ref{fig:decomps} are indeed well-captured by the shapelet models.

\subsection{Astrometry} \label{astrometry}

It can similarly be shown that the \com{unweighted} centroid ($x_c,y_c$) is

\begin{equation} \label{eqn:xc}
x_{c}+i y_{c} \equiv \frac{\iint_{\mathbb{R}} (x+iy) f({\mathbf x}) ~{\mathrm d}^{2}x}{\iint_{\mathbb{R}}
f({\mathbf x}) ~{\mathrm d}^{2}x} = \frac{
(8\pi)^{\frac{1}{2}} \beta^{2} }{F}~ \sum_{n}^{\rm odd}
(n+1)^{\frac{1}{2}} ~ f_{n1} 
\end{equation}

\noindent Here the summation is over only odd values of $n$, because only these
have the $m=\pm1$ coefficients that possess the desired rotational symmetries.

\subsection{Size}

Measures for the size and ellipticity of an object can be derived
from its unweighted quadrupole moments,

\begin{equation}
\label{eqn:f_ij} F_{ij} \equiv \iint_{\mathbb{R}} x_{i}x_{j} f({\mathbf
x})~{\mathrm d}^{2}x ~.
\end{equation}

\noindent The rms radius $R$ of an object is given by

\begin{eqnarray} \label{eqn:r2def}
  R^{2} & \equiv & \frac{\iint_{\mathbb{R}} |{\mathbf x}|^{2} f({\mathbf x})~{\mathrm d}^{2}x}
                      {\iint_{\mathbb{R}} f({\mathbf x})~{\mathrm d}^{2}x} \\
  & = & \frac{F_{11}+F_{22}}{F}
  ~ = ~~~ \frac{ (16 \pi)^{\frac{1}{2}} \beta^{3} }{F}~ \sum_{n}^{\rm even}
(n+1) ~ f_{n0}~. \label{eqn:r2}
\end{eqnarray}

\noindent Integrals (\ref{eqn:cs1}) to (\ref{eqn:int0rchinclosed}) can also be
used to calculate Petrosian radii that enclose a specified fraction of the total
flux within a circular aperture.

\subsection{Ellipticity}

The unweighted ellipticity of an object can also be calculated from its
quadrupole moments. 

\begin{equation}
\label{eqn:epsilon} \varepsilon \equiv
\frac{F_{11}-F_{22}+2iF_{12}}{F_{11}+F_{22}} =
\frac{(16\pi)^{\frac{1}{2}}\beta^{3}}{FR^{2}}
  \sum_{n}^{\rm even} [n(n+2)]^{\frac{1}{2}} f_{n2} ~,
\end{equation}

\noindent where the complex ellipticity notation of Blandford \etal\ (1991),
with $\varepsilon=|\varepsilon|\cos{2\theta}+i|\varepsilon|\sin{2\theta}$,
arises here naturally.

%
%
%
%

\section{Galaxy morphology classification} \label{hosm}

The shapelet decomposition of an object captures its entire structure, and
useful information is frequently found in coefficients of higher order than
those considered above. In particular, galaxy morphologies are well known to
provide an indication of their physical properties, local environment and
formation history. The classical ``Hubble sequence'' of morphological types has
been recently improved by several shape estimators that attempt to classify
galaxies in a more quantitative manner, which correlates directly with the
physical properties of interest ({\it e.g.}\ Simard 1998; Bershady, Jangren \&
Conselice 2000; van den Bergh 2002). 

It is possible to manufacure such morphology diagnostics from weighted
combinations of shapelet coefficients. Introducing shapelets to this field
allows a measurement to take advantage of our robust treatment of noise,
pixellisation and PSF deconvolution. The shapelet expressions for existing shape
measures are frequently elegant; and the natural symmetries in shapelets also
suggest new diagnostics.

\subsection{General scale-invariant quantities}

One approach is to consider general shape estimators $Q$, formed from a linear
combination of shapelet coefficients, as an extension of the
previous section

\begin{equation}
\label{eqn:qms}
Q=\beta^s~\sum_{n,m}w_{n,m}f_{n,m} ~,
\end{equation}

\noindent where $w_{n,m}$ are arbitrary weights, and the exponent $s$
sets the dimension of the estimator. These are also linear in a
galaxy's surface brightness. We initially restrict ourselves to using
those combinations which are independent of $\beta$ to at least first
order. This ensures that the choice of the scale factor does not
affect the final result, and is also equivalent to invariance under
object dilations (\ref{eqn:opdilateflux}). We can then impose further
constraints that the estimator must be independent of or linearly
dependent upon the various other operations described in
\S\ref{transformations}. Setting $\frac{\partial Q}{\partial
\beta}=0$ and using the result that

\begin{eqnarray}
\frac{\partial f_{n,m}}{\partial \beta} = \frac{1}{2\beta}
\left\{\sqrt{(n+m+2)(n-m+2)}~f_{n+2,m}\right. ~~~~~~ \nonumber \\
\left.-~\sqrt{(n+m)(n-m)}~f_{n-2,m}\right\}~,
\end{eqnarray}

\noindent it is easy to show that we require

\begin{eqnarray}
w_{n,m} & = & \frac{2s}{\sqrt{(n+m)(n-m)}}~w_{n-2,m} \nonumber \\
        &   & +~\sqrt{\frac{(n+m-2)(n-m-2)}{(n+m)(n-m)}}~w_{n-4,m} ~.
\end{eqnarray}

\noindent Notice that all quantities so formed mix coefficients with
only one value of $|m|$. This can be chosen to give $Q$ the desired
properties under rotation. Any term on the right-hand side should be
ignored if it refers to non-existent states with negative $n$. The
normalisation of the first term in each series, $w_{m,m}$, is
arbitrary: this can be set to ensure independence to changes of
object flux. 

Setting $(s,m)=(1,0), (2,1), (3,0)$ and $(3,2)$ recovers the flux $F$, centroid
$x_{c}$, rms square radius $R^{2}$ and ellipticity $\varepsilon$, up to the
normalisation factor of $F^{-1}$ for the latter three quantities. This proves
that these are indeed the {\it only} $\beta$-invariant linear quantities with
such dimensionality \com{and rotational symmetries}. Furthermore, since
equations~(\ref{eqn:f}), (\ref{eqn:xc}), (\ref{eqn:r2}) and (\ref{eqn:epsilon})
describe for {\it unweighted} moments, they must in fact be independent of
$\beta$ to all orders.

All of these basic shape estimators converge for any galaxy with a shapelet
spectrum steeper than $n^{-2}$. This includes both spiral galaxies with an
exponential profile, and elliptical galaxies with a ``de Vaucouleurs'' profile,
as long as $n_{\rm max}$ is kept sufficiently low to prevent the high-$n$
coefficients from modelling background noise at large radii. The flux and
centroid estimators converge most rapidly, so are least sensitive to the choice
of $n_{\rm max}$. The error on these series due to truncation can be calculated
using any of a range of methods for generic Taylor series in {\it e.g.}\ Boas
(1983).

\subsection{Concentration}

We can extend this sequence by raising $s$ further. For example,
setting $s=5$ and $m=0$ gives the 2D unweighted kurtosis of the
image, producing an estimate of the object's concentration.
Unfortunately, such a high value of $s$ yields a series of shapelet
coefficients that does not converge for galaxies with a de
Vaucouleurs or exponential radial profile.

We have also noticed that a ratio of the two existing shapelet scale sizes, $R$
and $\beta$, also works rather well as a concentration index (although this
estimator is not independent of $\beta$). Further work will need to be done to
calibrate this estimator to the physical properties of galaxies.

An alternative approach is to mimic pre-existing, and pre-calibrated, morphology
diagnostics. Bershady, Jangren \& Conselice (2000) define a concentration index

\begin{equation} \label{eqn:c}
  C \equiv 5 \times \log \left( \frac{r_{80}}{r_{20}} \right) ~,
\end{equation}

\noindent where $r_{80}$ and $r_{20}$ are the radii of circular apertures
containing 80\% and 20\% of the object's total flux. This correlates well with a
galaxy's Hubble type (Bershady, Jangren \& Conselice 2000) and also its mass
(Conselice, Gallagher, \& Wyse 2002). Integrals (\ref{eqn:cs1}) to
(\ref{eqn:int0rchinclosed}) can be evaluated for various values of $R$, to find
$r_{80}$ and $r_{20}$, and thus calculate this quantity for a shapelet model.

\subsection{Asymmetry}

Conselice, Bershady \& Jangren (2000) define an asymmetry index

\begin{equation} \label{eqn:a180}
A\equiv
 \frac{\sum_{\rm pixels}|f(x,y)-f^{180^\circ}(x,y)|}
      {\sum_{\rm pixels}f(x,y)}  ~,
\end{equation}

\noindent where the superscript denotes an image rotated through $180^\circ$. A
term dealing with the background noise and sky level has been omitted here, as
these are automatically dealt with during the shapelet decomposition process in
\S\ref{mk_decomp}. Asymmetry correlates with a galaxy's star formation rate
(Conselice, Bershady \& Jangren 2000), and high asymmetry values often indicate
recent galaxy interactions or mergers.

In a shapelet expansion, all of the information about galaxy asymmetry is
contained in coefficients with odd $m$ (and $n$). Using the orthonormality
condition (\ref{eqn:orthop}) and rotating via equation~(\ref{eqn:oprot180}), we
find the simple form

\begin{equation} \label{eqn:curv}
A=\frac{\sqrt{2}\beta}{\pi F}~\sum_{n,m}^{\mathrm odd}|f_{n,m}| ~.
\end{equation}

Estimators of asymmetry under rotations of $120^\circ$ or $90^\circ$ can also be
formed from sums of shapelet coefficients with $m$ not divisible by 3 or 4
respectively.

\subsection{Chirality} \label{chirality}

A quadratic combination of shapelet coefficients can be used to describe the
``chirality'' or ``handedness'' of an object. One dimensionless estimator
$\chi_{|m|}$ can be formed for each value of $|m|$, to trace the relative
rotation of those coefficients, with increasing $n$. This is roughly equivalent
to tracing the rotation of a galaxy's isophotes with increasing radius. For
example, the galaxy shown in figure~\ref{fig:hstgal} has two prominent spiral
arms that unwind in a clockwise sense, so it has a large, positive value of
$\chi_2$.

We require that the chirality estimators should be invariant under global
rotation of the object; invariant under changes of flux; invariant to first
order under changes of $\beta$; and to flip sign when the object is
mirror-imaged. These conditions uniquely specify

\begin{equation}
\chi_{|m|} = \frac{\beta^2}{F^2}
             \sum_{n=m}^\infty \sum_{n'=n+2}^\infty
             w_{n,n'} f_{n,m}^* f_{n',m} ~,
\end{equation}

\noindent where $w_{m,m+2}=1$ and 

\begin{equation}
\sqrt{(n'+m)(n'-m)}w_{n,n'} = 4w_{n,n'-2} + \sqrt{(n+m)(n-m)} ~,
\end{equation}

\noindent thus mixing all coefficients with the same value of $m$.

This estimator has yet to be calibrated against physical quantities. However,
this approach ought to be able to automatically distinguish between elliptical
galaxies and spiral galaxies in way that mimics a visual classification, and
could also be adapted as a function of $n_{\rm max}$ to find bars in the cores
of spiral galaxies.

\section{Conclusions} \label{conc}

We have extended the formalism of shapelets for image analysis from basis
functions separable in Cartesian to polar coordinates. Cartesian shapelets are
convenient for the initial object decomposition. In particular, we have shown
that that they can analytically be integrated inside a square boundary, thus
facilitating the pixellisation of the smooth basis functions. On the other hand,
polar shapelets decompose an object into components with explicit rotational
symmetries, and often have a more direct physical interpretation. In addition,
they yield more compact representations of typical galaxy images, since terms
with low orders of rotational symmetry tend to dominate.

We have quantitatively investigated the effects of the choice of the shapelet
scale size parameter, $\beta$. For most objects in astronomical images, one
scale size is clearly optimal for high quality image reconstruction, data
compression and the fast convergence of shape estimators. We have developed a
practical algorithm to find this value of $\beta$ for arbitrary objects in real
images, plus optimum values for the shapelet centre ${\mathbf x}_c$ and
truncation order $n_{\rm max}$. This algorithm can also take into account
observational effects including noise, pixellisation and PSF deconvolution.

We have then described a number of applications of polar shapelets. Shapelet
models can be rotated, enlarged and sheared by simple analytic operations. Since
the shapelet basis functions are invariant under Fourier transform, analytic
convolutions and deconvolutions ({\it e.g.}\ from a PSF) are also easy to
perform. Linear combinations of an object's polar shapelet coefficients produce
elegant expressions for its flux (photometry), position (astrometry) and size.
We also showed how other combinations of shapelet coefficients can be used to
distinguish between morphological types of galaxies. A complete {\tt IDL}
software package to perform the image decomposition and shapelet image analysis
is publicly available on the world wide web at {\tt
http://www.astro.caltech.edu/$\sim$rjm/ shapelets/}.



\section*{Acknowledgments}

The authors thank David Bacon, Gary Bernstein, Sarah Bridle, Tzu-Ching Chang,
Mark Coffey, Chris Conselice, Phil Marshall and Jean-Luc Starck for invaluable
insights and constructive conversations. Astute suggestions from the referee
helped shape this paper into a more logical progression.

\appendix
\section{Laguerre polynomials}
\label{recursion}

Different conventions have been used to define the Laguerre polynomials,
especially before the 1960s. The $p!$ term is omitted from
equation~(\ref{eqn:polardef}) in many older books, and caution must be observed
with the resulting relations. Several useful recursion relations can be derived
to simplify their calculation ({\it e.g.}~Boas 1983, Ch.12), which we gather
here, using our convention, for future reference:

\begin{eqnarray} 
L^q_0(x)~~~ & = & 1 \label{eqn:laguerrepoly0} \\
L^q_1(x)~~~ & = & 1 - x + q  \label{eqn:laguerrepoly1}\\
L^q_p(x)~~~ & = & \left(2+\frac{q-1-x}{p}\right)L^q_{p-1}(x) \nonumber \\
            & ~ & ~~~~~~~~~~~~~~~~~~~~~~~~-\left(1+\frac{q-1}{p}\right)
                  L^q_{p-2}(x) \label{eqn:laguerrerecursion1} \\
            & = & L^{q+1}_p(x)-L^{q+1}_{p-1}(x) \\
            & = & \sum_{i=0}^pL^{q-1}_i \\
\frac{{\mathrm d}L^q_p(x)}{{\mathrm d}x}
            & = & x^{-1} \left\{pL^q_p(x)-(p+q)L^q_{p-1}(x)\right\} \\
            & = & -L^{q+1}_{p-1}(x) \label{eqn:laguerrerecursion3} \\
\frac{{\mathrm d}L^0_{p}(x)}{{\mathrm d}x} 
            & = & \frac{{\mathrm d}L^0_{p-1}(x)}{{\mathrm d}x} - L_{p-1}^0(x)
	          \label{eqn:laguerrerecursion4} \\
            & = & -\sum_{i=0}^{p-1}L^0_i~ \label{eqn:laguerrerecursion2}.
\end{eqnarray}

\bsp
\label{lastpage}

\end{document}